\RequirePackage{ifpdf} 

\documentclass{pos}

\title{Recent results of the CMS experiment}

\ShortTitle{Recent results of the CMS experiment}

\author{
        Lars Sonnenschein$^a$ \\ 
        \scalebox{1.2}{\rm on behalf of the CMS collaboration} \\ \\
        $^a$III. Phys. Inst. A \\
        RWTH Aachen University\\
        E-mail: \email{Lars.Sonnenschein@cern.ch}\\
       }

\abstract{The CMS experiment is a multi-purpose detector successfully operated at the LHC
where predominantly $pp$ collisions take place at various centre-of-mass energies
up to $\sqrt{s}=8$~TeV so far. Several weeks per year also heavy-ion collisions 
take place leading to interesting studies in $Pb$-$Pb$ and $p$-$Pb$ collisions
at $\sqrt{s_{NN}}=2.76$~TeV and $\sqrt{s_{NN}}=5.02$~TeV centre-of-mass energies per nucleon,
respectively. The excellent performance of the accelerator 
and the experiment allows for dedicated physics measurements over a wide range
of subjects, starting from particle identification, encompassing 
forward physics, Standard Model measurements in multijet, boson, heavy flavour and top quark 
physics, building the basis for new physics searches interpreted within the framework
of various models and theories. These pursued $pp$ physics subjects are complemented by 
a rich heavy ion physics programme.} 

\FullConference{52nd International Winter Meeting on Nuclear Physics,\\
                January 27-31, 2014\\
                Bormio Italy}

\begin{document}

\section{Introduction} \label{intro}
The performance of the Compact Muon Solenoid (CMS) experiment~\cite{cmsdet} is discussed 
in detail, followed by measurements dedicated to give a more accurate account
of the Standard Model (SM) by means of forward physics,
multijet and boson production, followed by a rich $B$ physics programme 
setting severe constraints on new physics by means of the rare decays 
$B_{(d,s)}^0\rightarrow \mu^+\mu^-$. 
Furthermore the heaviest Standard Model candle - the top quark -
is produced copiously at the Large Hadron Collider (LHC), providing the opportunity 
for detailed studies
of $t\bar{t}$ and single top production cross sections as well as measurements 
of top quark properties, most prominently its mass.
The Higgs boson is discussed in various decay channels and the properties like its 
mass, signal strengths and couplings are elaborated.
The search for supersymmetry is exemplarily
elaborated for hadronic, semi-leptonic and leptonic final states, followed by 
a summary of excluded mass scales in simplified model mass spectra (SMS) and exclusion
limits in the phase space of the Constrained Minimal Supersymmetric SM (CMSSM).
Afterwards an exotic search is presented exemplarily.
Finally a few important heavy ion physics measurements are discussed.

\noindent
Throughout this article the convention $c \equiv 1$ is adopted for the speed of light.

\begin{figure}[b]
 \vspace*{-4ex}
 \resizebox{0.5\columnwidth}{!}{%
 \hspace*{-4ex} \includegraphics{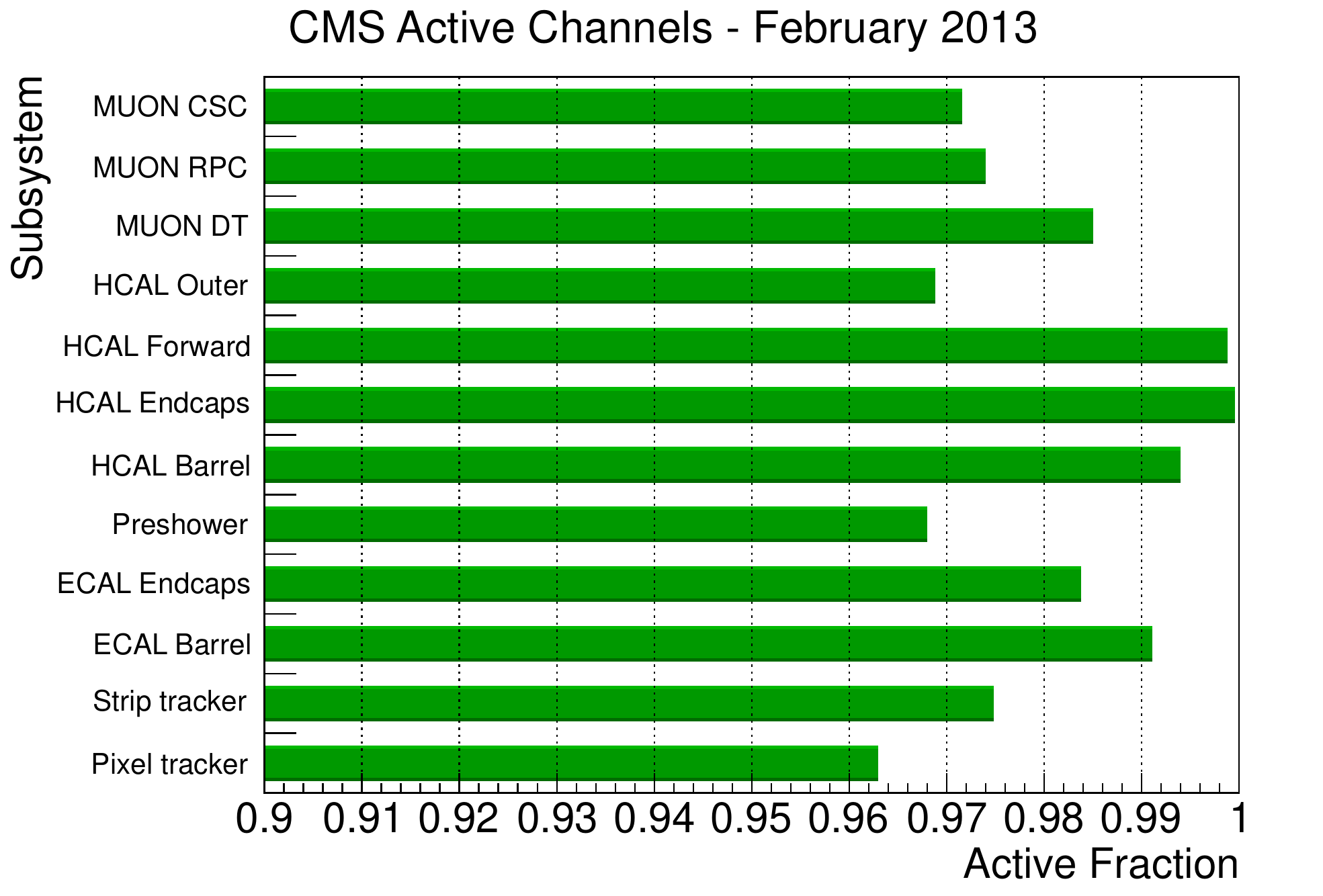} 
 \hspace*{-4ex}
}
 \unitlength 1cm
 \begin{picture}(10.,6.)
 \put(-0.4, 5.6){ \includegraphics[width=6.0cm, angle=270]{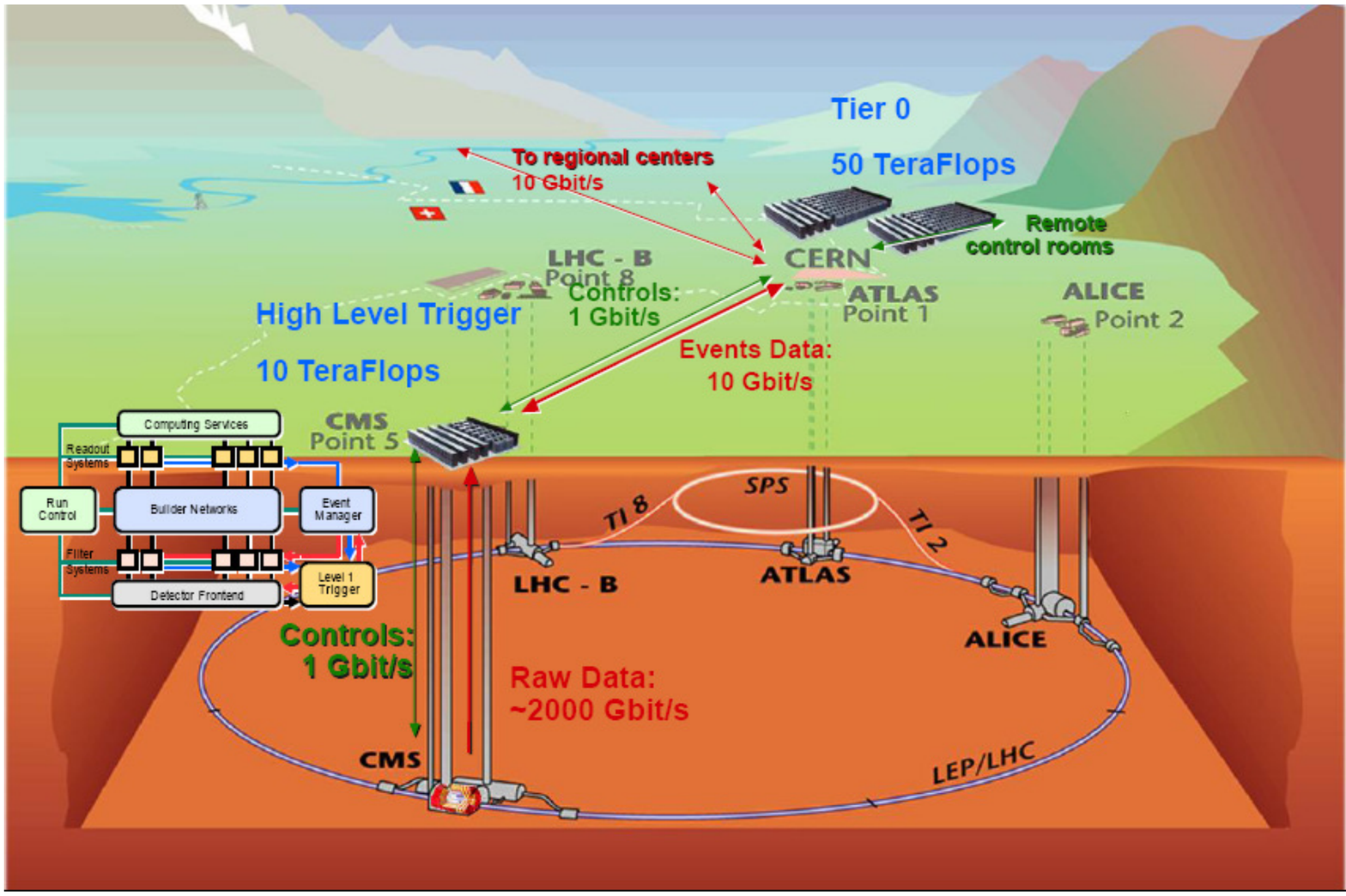} }
 \put(7.25, 0.7){\rotatebox{90}{\scalebox{0.4}{Credit: C. Schwick/CERN}}}
\end{picture}

\caption{ \label{CMSperformance}
Active fraction of the CMS subdetectors (left). In the average 98\% of the 
detector have been operational before going into the Long Shutdown 1 (LS1)
maintenance phase in which repairs, optimisations and upgrades take place. 
On the right is shown an overview of the LHC experiments, 
in particular the CMS experiment with its infrastructure, trigger system and data 
flow.
}
\end{figure}

\section{CMS status and operation}
The Compact Muon Solenoid (CMS) experiment~\cite{cmsdet} is a multi-purpose 
detector operated a the Large Hadron Collider (LHC) at CERN. In the barrel part 
of the detector concentric around the beam pipe are installed from inside to 
outside three layers of silicon pixels (66 mio. pixels),
surrounded by ten layers of silicon strips (adding 9.6 mio. channels). 
This inner tracking system is surrounded by the
electromagnetic crystal calorimeter consisting of 76~000 scintillating $PbWO_4$ mono-crystals and the hadronic brass sampling calorimeter consisting of brass as absorber material, equipped with plastic scintillators in slits for the sampled readout, summing up to about 7~000 channels.
A superconducting solenoid with an inner homogeneous magnetic field of 3.8~T is 
surrounding the hadronic calorimeter. Its purpose is to bend the charged particle 
tracks, which in turn allows the determination of their momenta once the particle 
type is identified. 
Outside of the magnet are located four layers of gaseous muon chambers
with excellent spatial resolution, complemented by resistive plate chambers
with excellent temporal resolution, 
embedded between the
iron return yoke to focus the magnetic flux inside the detector volume.
The magnetic field at the muon detectors outside the solenoid
amounts in average to about 2~T.
Similar detector technologies are extended into the endcaps in form of disks
for hermetic coverage.

The different subsystems of the CMS detector are operational to 98\% in average 
(see Fig.~\ref{CMSperformance}, left), ranging from
the highest data taking efficiency of nearly 100\%
reached by the hadronic calorimeter, to the lowest data taking efficiency
of 96.3\% of the silicon strips. Altogether there are bout 80 million channels to be read 
out. During the Long Shutdown 1 (LS1) which is ongoing since February 2013 defect 
channels are being recovered. The high efficiency helps to keep the reconstruction and 
the comparison to simulation simple.

The rate of $pp$ collisions produces a data volume corresponding to about 2~TB per second. To reduce this huge amount of data to a manageable size, a Level-1 trigger based on fast electronics implemented in programmable chips is employed. With an acceptance 
rate of up to 100~kHz the data is then send to a computing farm at CMS
with 2400 multicore Linux PC's, where the events are scrutinised with more 
sophisticated algorithms to make a final decision at this 
High Level Trigger~\cite{triggerTwiki} (HLT),
if a given event might be of interest or can be discarded. The HLT accept rate
can reach values closely above 2000~Hz. The rate of the most important triggers 
at HLT level agree with simulation within 5\%.
Over 300~TB local disk space permits the 
temporary storage of the data for up to one week. In normal operation the data is send
continuously to the CERN computing centre for storage, reconstruction and distribution to computing centres of participating institutions around the world.
An overview of the CMS infrastructure and data flow is 
depicted in Fig.~\ref{CMSperformance}, right.

In 2011 (2012) up to 130~pb$^{-1}$ (286~pb$^{-1}$) of integrated luminosity per day have 
been \linebreak 
recorded~\cite{lumiPAS}\cite{lumiTwiki}.
Peak luminosities up to $40 \cdot 10^{32}$~Hz/cm$^2$ ($77 \cdot 10^{32}$~Hz/cm$^2$) have been reached in 2011 (2012) and 
integrated luminosities of 6.1~fb$^{-1}$ (23.3~fb$^{-1}$) have been delivered by the LHC
which had a proton run efficiency of 36.5\% in 2012. The remainder of the time is 
dedicated to SetUp, Injection, Access, Squeeze and Ramp periods in decreasing order. 
The recorded data undergoes a certification process where depending on the
operationality of the various sub-detector systems and the needs for dedicated physics 
analyses two dataset categories are established. 85\% of the data are certified as good
where all sub-detectors have been fully operational. 90\% of the data are certified
as good for muon analyses disregarding calorimeter quality.

The number of parasitic $pp$ collisions (referred to as pile-up events) 
accompanying a high energetic scattering of 
interest in a single LHC beam crossing is distributed according to a Poisson statistics 
with an average number of nine to ten events in 2011 and up to 25 events in 
2012. In general the distinct vertices can be reconstructed to sort out interesting 
high energetic physics events from additional low energetic scattering background.
The number of pile-up events increases with the number of instantaneous luminosity.
A compromise is made between high luminosity and clean events for an optimised 
discovery potential.

In the last LHC run period (Run II) until LS1 proton proton collisions took place at a 
centre-of-mass energy of $\sqrt{s}=8$~TeV with a proton bunch spacing of 50~ns. 
Data at previous centre-of-mass energies of $\sqrt{s}=0.9$, 2.76 and
7~TeV with the same bunch spacing have been analysed and are partly presented here. 
Dedicated run periods with heavy ion collisions
at a centre-of-mass energy of 2.75~TeV (5.02~TeV) per nucleon in $Pb$-$Pb$ ($p$-$Pb$) 
collisions do also take place.
One of the primary goals why the LHC has been built is the discovery of new physics,
in particular the electro-weak symmetry breaking mechanism, also referred to as 
Brout-Englert-Higgs (BEH) mechanism which has been primarily conceived to endow 
the particles with mass. 
Experimental analysis of the Higgs boson particle predicted by this mechanism permits 
to probe the mathematical consistency
of the Standard Model (SM) of particle physics beyond the TeV energy scale.
The exploration of new energy frontiers at collider based experiments opens new possibilities
on the way toward a unified theory. Potential discoveries could reveal the existence of
supersymmetry, extra dimensions or dark matter at accessible energy scales.
Parallel to the searchs for new particles and new phenomena a rich programme of SM physics
in various domains are conducted, including forward physics, multijet production, 
electro-weak vector boson production, heavy flavour physics, top quark physics. 
Furthermore heavy-ion collisions are exploited among others to study dense matter phases 
like quark-gluon plasma.

\begin{figure}[b]
 \vspace*{-3ex}
 \resizebox{0.46\columnwidth}{!}{%
 \hspace*{-4ex} \includegraphics{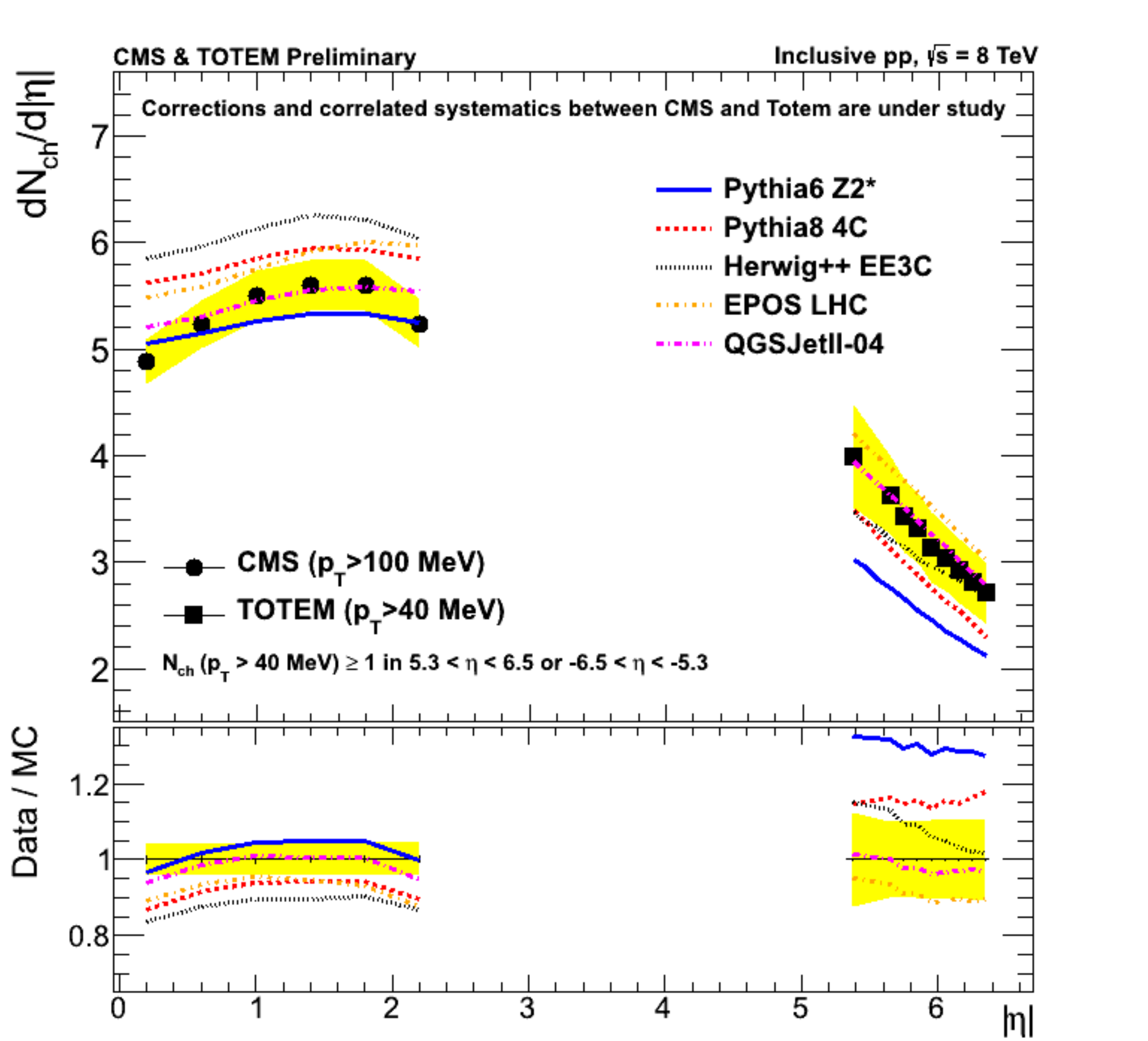} 
 \hspace*{-4ex}
}
 \unitlength 1cm
 \begin{picture}(10.,6.)
 \put(-0.45, -0.25){ \includegraphics[width=8.6cm]{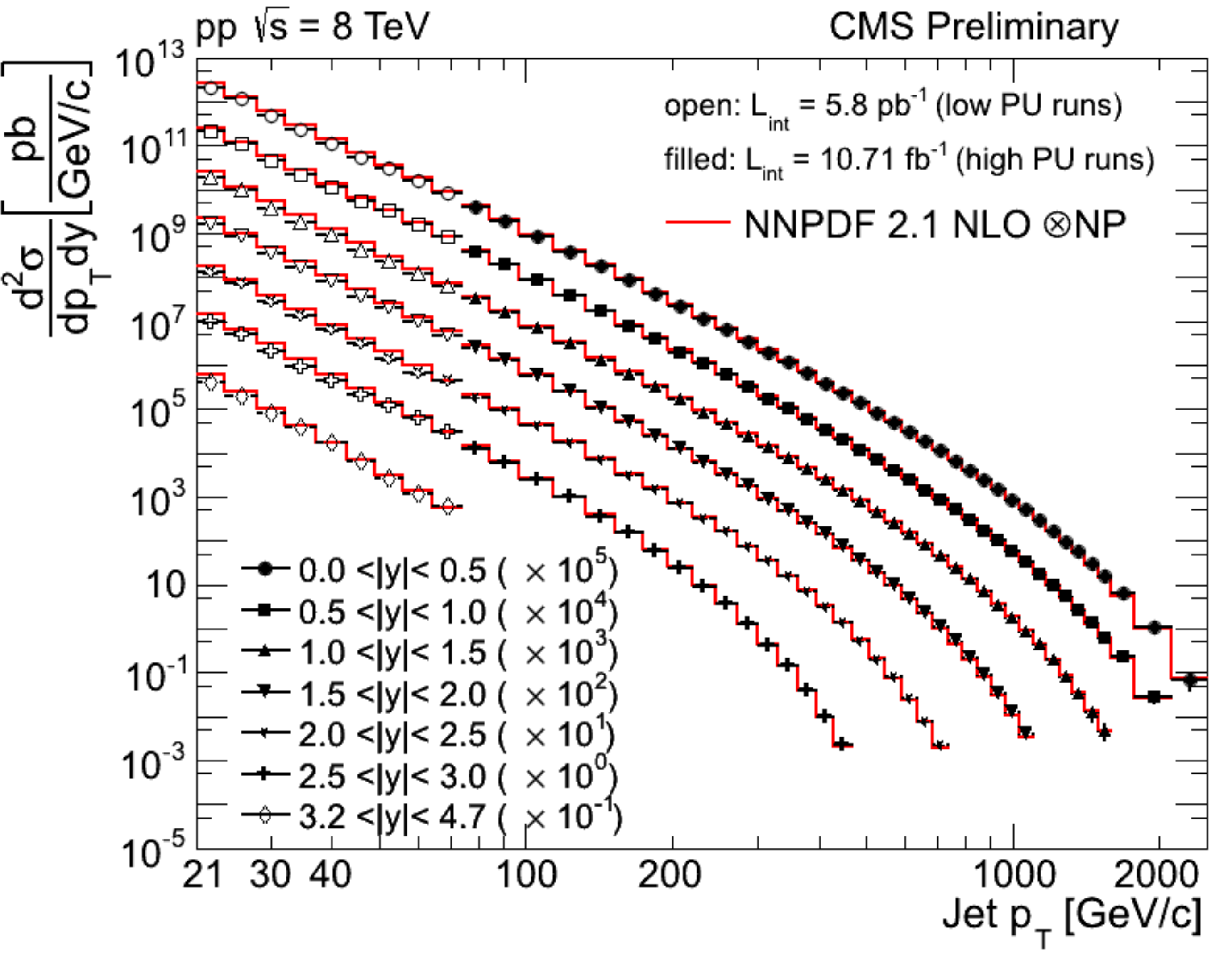} }
\end{picture}

\vspace*{-1ex}

\caption{ \label{ForwardPlots}
Left: Charged particle multiplicity distribution as a function of $|\eta|$. CMS
covers the central pseudorapidities $|\eta|<2.4$ and TOTEM the forward range $5.3<|\eta|<6.5$.
Right: Inclusive jet cross section differentially in jet transverse momentum $p_T$ and rapidity
$y$. The low momentum region $21<p_T<74$ GeV is covered by low pile-up runs.
The high momentum region $p_T>74$ GeV is covered by high pile-up runs.
}
\end{figure}

\section{Forward and low $x$ scale physics}
At the LHC a wide range of proton momentum fractions $x$ and momentum scales $Q$ are accessible, 
extending the phase space of previous colliders considerably. In particular measurements at 
small $x$ and scale are of interest for airshower experiments like Auger 
to improve the understanding and description of the interactions of cosmic rayons in the 
atmosphere. Small $x$ at the LHC implies forward proton and diffractive physics while
low scale implies soft physics in the underlying event. Several diffractive jet 
production mechanisms can be distinguished. 
In single diffractive $pp$ collisions, where the two 
protons interact via one colourless double gluon exchange, also referred to as Pomeron,
a rapidity gap separates the central hard scattering of the resolved proton and the proton which 
leaves the interaction intact into the forward direction. 
In double diffractive $pp$ collisions both protons which interact via a Pomeron get resolved
and a rapidity gap is expected in the central region between the two hard jet activity regions.
A further possibility provides the double Pomeron exchange, where both protons leave the 
collision intact into the forward direction but separated by two rapidity gaps 
a hard jet scattering takes place in the central rapidity region.
In the following some forward and low $x$ momentum fraction measurements are discussed.

The inelastic $p-Pb$ cross section at a nucleon nucleon centre-of-mass energy of 
$\sqrt{s_{NN}}=5.02$ TeV has been measured~\cite{pas-fsq-13-006}.
Inelastic collisions are tagged using the hadronic forward calorimeters at pseudorapidities
$3<|\eta|<5$. Two different event selections are used: A coincidence in both sides
of the hadronic forward calorimeters and a single-sided event selection. 
These two selections have different sensitivity to contributions from photon induced 
($\gamma p$) collisions and hadronic diffractive interactions. 
The value of the hadronic inelastic cross section is measured
within the CMS acceptance and extrapolated to its total value. The photon-induced
contribution is subtracted. The final result $\sigma_{\mbox{\scriptsize inel.}}=2.06 \pm 0.08$ b. 
The uncertainty is dominated by the determination of the luminosity. This measurement of the 
inelastic cross section is consistent with the expectation from the Glauber~\cite{glauber} 
approach ($\sigma_{\mbox{\scriptsize inel.}} = 2.13 \pm 0.04$ b), based on the $pp$ cross section 
from the COMPETE~\cite{compete} fit that has been extended by the 
TOTEM~\cite{totem} measurement at a centre-of-mass energy of 7 TeV.
Other model predictions are also compatible with the measurements given the experimental 
uncertainties. Only QGSJETII-04 predicts a value slightly above the upper uncertainty bound of 
this measurement.

\begin{figure}[b]
 \vspace*{-3ex}
 \resizebox{0.47\columnwidth}{!}{%
 \hspace*{-4ex} \includegraphics{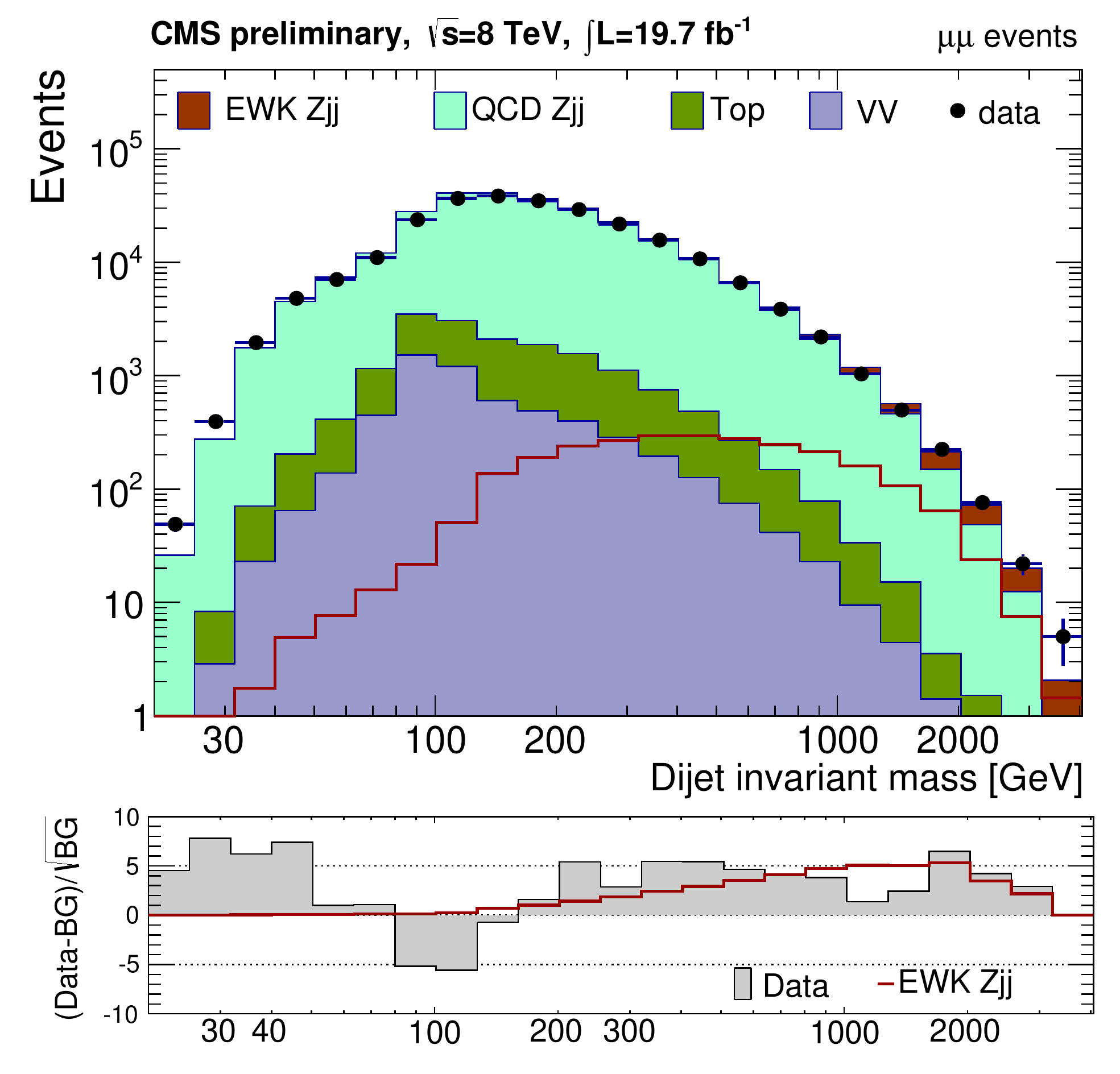} 
 \hspace*{-4ex}
}
 \unitlength 1cm
 \begin{picture}(10.,6.)
 \put(0.1, -0.3){ \includegraphics[width=7.7cm]{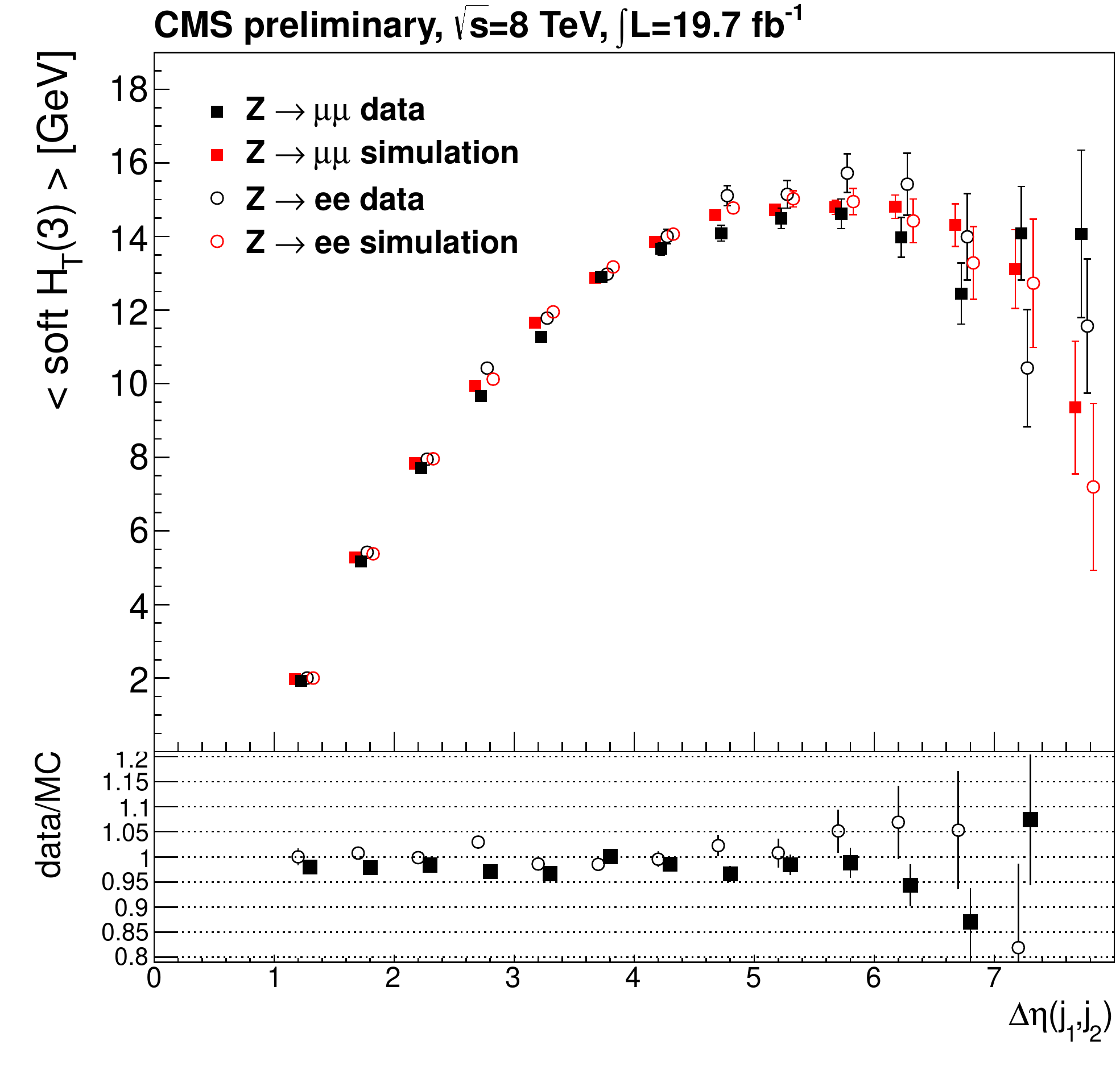} }
\end{picture}

\caption{ \label{ZjjDistributions}
Dijet invariant mass distribution. The data is compared to different background contributions 
dominated by QCD production of $Zjj$ and the signal (left).
The scalar sum of the three leading transverse momentum track jets 
as a function of the pseudorapidity difference between the two tagged jets is shown in the right
plot. The data 
is indicated in black, the simulation in red. The Drell-Yan (DY) dimuon channel is presented 
in squares, the dielectron channel in circles.}
\end{figure}

Pseudorapidity and leading transverse momentum distributions of charged particles in $pp$ 
collisions at $\sqrt{s} = 8$ TeV have been measured with the CMS and TOTEM 
detectors~\cite{pas-fsq-12-026} in the pseudorapidity range of $|\eta| < 2.4$.
The data correspond to an integrated luminosity of 17.4 nb$^{-1}$. Events are triggered by the 
TOTEM T2 telescopes covering the pseudorapidity range of $5.3 < |\eta| < 6.5$ for reconstructed 
tracks above a transverse momentum threshold of 40 MeV.
The measurement of the pseudorapidity distributions is performed for primary
charged particles with $p_T > 0.1$ GeV and $p_T > 1$ GeV for two different selection criteria.
An inclusive sample is obtained by requiring tracks reconstructed in the acceptance of
the TOTEM T2 telescopes in either the forward or the backward hemispheres. 
The differential charged particle distribution as a function of $|\eta|$ is shown in
Fig.~\ref{ForwardPlots} (left), covering the central CMS and forward TOTEM measurements,
obtained by making use of low pile-up runs.
Another sample enhanced in non-single diffractive dissociation events is obtained by 
requiring tracks in the T2 telescope in both forward and backward hemispheres. 
The average multiplicity per unit of pseudorapidity has been measured to be $5.4 \pm 0.2$ for the 
inclusive selection and $6.2 \pm 0.3$ for the Non Single Diffractive (NSD) enhanced sample, 
while for $p_T > 1$ GeV it has been measured to be $0.78 \pm 0.03$ and $0.93 \pm 0.04$,
respectively.
At $\sqrt{s}=8$ TeV, the average value of $\left.dN_{ch}/d\eta\right|_{\eta =0}$ for the 
NSD-enhanced sample follows a power-like centre-of-mass energy dependence already 
suggested by previous NSD measurements at different energies. 
The measured charged-particle distributions can help to constrain the modelling of 
semi-hard (multi-)parton scattering processes in $pp$ collisions at
the LHC over large phase space regions in $p_T$ and $|\eta|$.
The transverse momentum distribution of the leading tracks in the central region is 
measured, too. The distribution integrated over the leading track transverse momentum above 
a $p_{T,\min}$ threshold shows a transition from a steeply falling distribution at large
$p_T$ (the perturbative region) to a flat distribution at small $p_T$ 
(the non-perturbative region) for $p_{T,\min}$ of a few GeV. This region is not well described 
by current Monte Carlo event generators.

The inclusive jet cross section in proton-proton collisions at $\sqrt{s} = 8$ TeV has been 
measured~\cite{pas-fsq-12-031} differentially in $p_T$ for seven rapidity bins based on data 
collected with the CMS detector. The data set corresponds to an 
integrated luminosity of $5.8$ pb$^{-1}$. The measurement was performed using the 
anti-$k_T$ cluster jet algorithm with a distance parameter of $R = 0.7$ in the rapidity region 
of $|y| < 4.7$ for jet transverse momenta of $21 < p_T < 74$ GeV. 
The measured differential cross section is shown in Fig.~\ref{ForwardPlots} (right) in 
comparison to the NLO theory prediction. The open points cover this low momentum measurement.
The dominant source of systematic uncertainty arises from the Jet Energy Scale (JES) which 
propagates about 5 - 45\% uncertainty on the cross section measurement across varying rapidity 
bins. The measured jet cross-section is compared to predictions of next-to-leading order (NLO) 
QCD calculations taking non perturbative (NP) and Parton Shower (PS) forward folding into account.
Within the current experimental and theoretical uncertainties, perturbative QCD calculations are 
in agreement with the measured inclusive jet cross sections.

The electro-weak production cross section of $Z$-bosons in the same flavour dilepton
final state associated with two forward/backward jets in proton-proton collisions at $\sqrt{s}=8$ 
TeV has been measured~\cite{pas-fsq-12-035} based on 
a data sample recorded by the CMS experiment at the LHC with an integrated luminosity of 
19.7 fb$^{-1}$. Different methods have been used to extract the signal,
including a data-driven approach that models the QCD $Zjj$ background based on
$\gamma jj$ events. The cross section for this process is measured in dielectron and dimuon 
final states in the kinematic region with invariant masses of $m_{\ell\ell} > 50$ GeV,
$m_{jj} > 120$ GeV, transverse momenta of $p_{Tj} > 25$ GeV and pseudorapidity $|\eta_j|<5$.
The dijet invariant mass distribution is given in Fig. \ref{ZjjDistributions}, left
and the distribution of the soft hadronic activity between the two jets is given at the right.
Combining two different methods and channels, the cross section has been measured to
$\sigma = 226 \pm 26(\mbox{stat}) \pm 35(\mbox{syst})$ fb, in agreement with the
theoretical cross section at NLO which amounts to 239 fb. 
The hadronic activity in events with $Z$-boson 
production in association with jets is also studied, in particular in the rapidity interval 
between the two jets. The measured gap fractions for soft-hadronic vetoes show good agreement
with different predictions.

\section{Standard Model physics}

The inclusive jet cross section, double-differential in jet transverse 
momentum $p_T$ and absolute jet rapidity $|y|$, has been measured~\cite{pas-smp-12-012}. 
Data from LHC
proton-proton collisions at $\sqrt{s} = 8$ TeV, corresponding to an integrated luminosity
of 10.71 fb$^{-1}$ have been collected with the CMS detector. Jets are reconstructed with
the anti-$k_T$ clustering jet algorithm with a distance parameter $R=0.7$ in a phase space 
region covering jet transverse momenta between 74 GeV and 2.5 TeV and an absolute rapidity
up to $|y|=3.0$. The parton momentum fractions $x$ probed in this measurement are given by 
the interval $0.019 < x < 0.625$.
The measured differential cross section is shown in Fig.~\ref{ForwardPlots} (right) in 
comparison to the NLO theory prediction. The filled points cover this high momentum measurement,
obtained by making use of high pile-up runs.

Dominant sources of experimental systematic uncertainties arise due to the jet energy scale 
(JES), unfolding and the luminosity measurement uncertainty. These lead to about 
15-40\% uncertainty in the cross section measurement across various rapidity bins. 
In comparison, the theory predictions are most affected by Parton Distribution Function (PDF) 
uncertainties, which amount to 10\% to 50\% depending on the rapidity bin, whereas choices of 
renormalisation and factorization scales contribute with 5\% to 10\% at central rapidity and 
with 40\% in the outer rapidity bins.
The measured jet cross section is corrected for detector effects and 
compared to predictions of perturbative QCD at NLO using various sets
of PDF's.
Some differences between the predictions employing various PDF sets and the data are 
observed, all of which are qualitatively covered by uncertainties in the low $p_T$ region, 
except for the ABM11 PDF set, which exhibits significant deviations. In the high $p_T$
region the CT10 theory prediction is in good agreement with data, whereas other PDF sets 
have significant differences compared to the measured cross section.
This new inclusive jet cross section measurement probes a wide range in $x$
and momentum scale $Q$. Therefore it can be exploited to constrain PDF's and to determine 
the strong coupling constant $\alpha_s$ in a new kinematic regime.

\begin{figure}[b]
 \vspace*{-4ex}
 \resizebox{0.47\columnwidth}{!}{%
 \hspace*{-4ex} \includegraphics{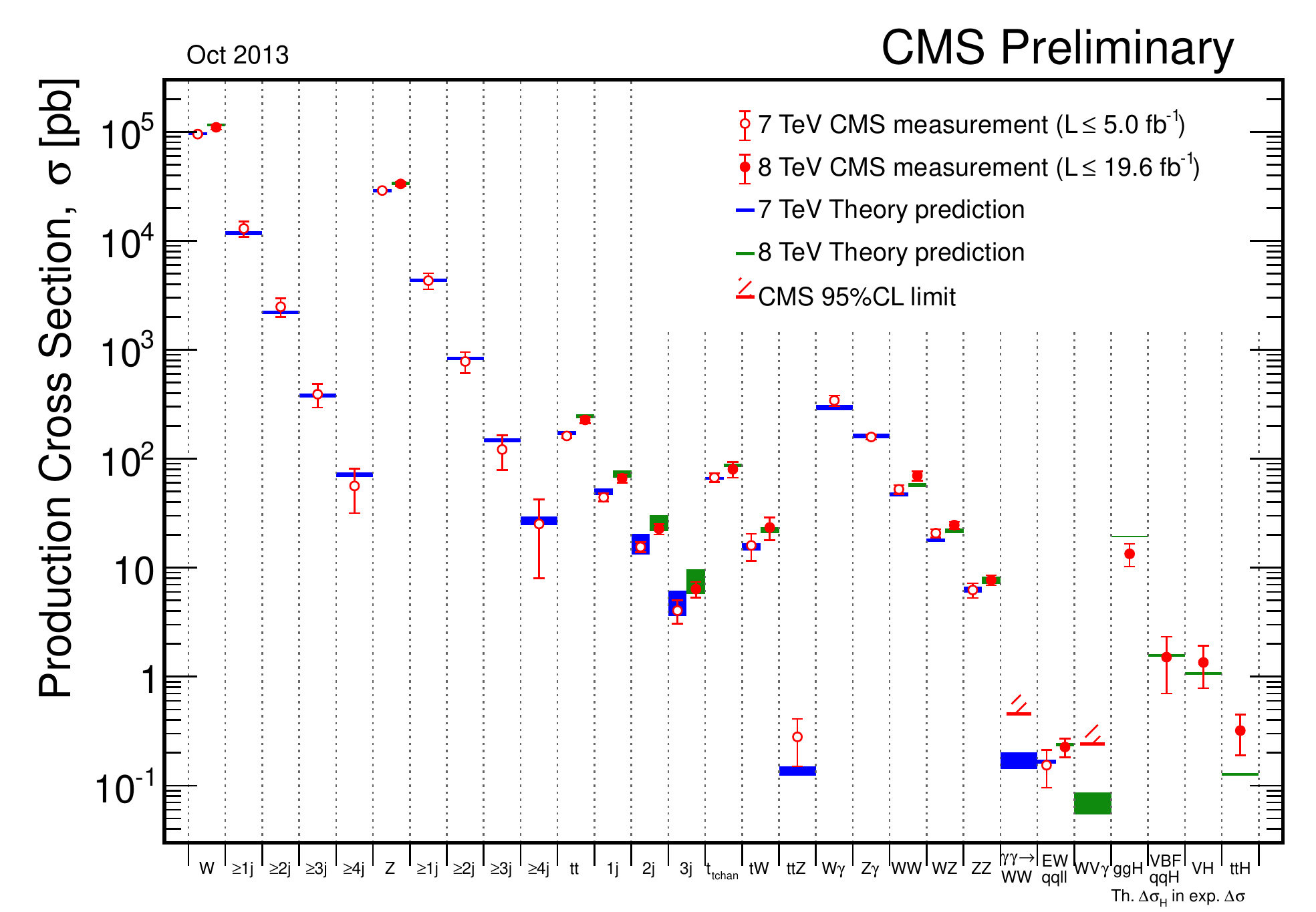} 
 \hspace*{-4ex}
}
 \unitlength 1cm
 \begin{picture}(10.,6.)
 \put(0.2, 0.0){ \includegraphics[width=7.7cm]{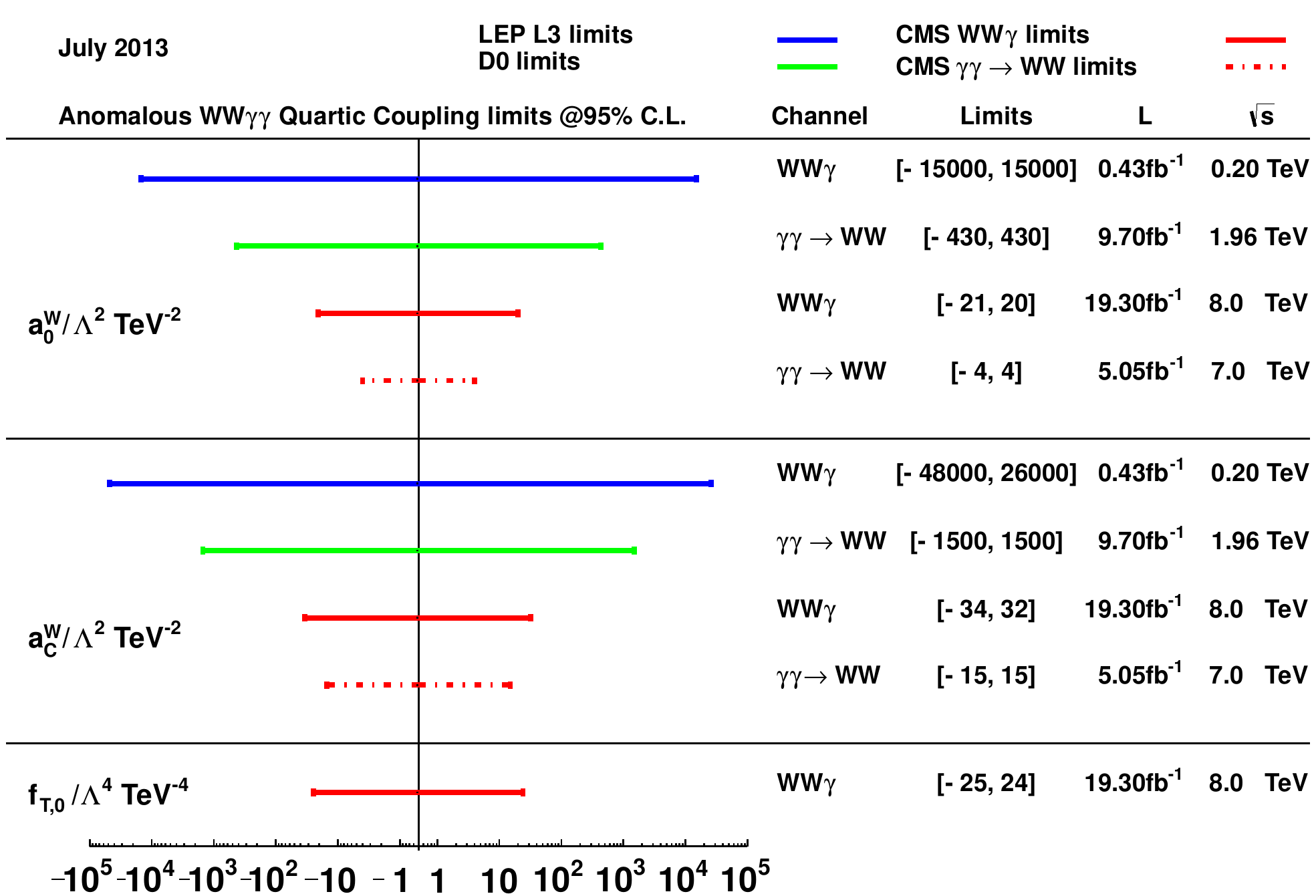} }
\end{picture}

\caption{ \label{SMprocesses}
Left: Standard Model production cross sections for various boson production processes
with and without additional jets for $\sqrt{s} =$7 and 8 TeV. CMS data points in red,
theory prediction lines in blue (7 TeV) and black (8 TeV). Right: Limits on quartic gauge 
couplings for the production of $WW\gamma$ and $\gamma\gamma\rightarrow WW$. CMS limits are 
indicated in comparison to L3 limits from LEP and D\O limits from Tevatron.}
\end{figure}
 
The inclusive jet cross section has been measured at $\sqrt{s} = 7$ TeV~\cite{pas-smp-12-028}
extending the accessible phase space in jet transverse momenta up to 2 TeV. The absolute jet 
rapidities range up to 2.5.
The experimental uncertainties of this measurement are smaller
than in previous publications such that these data constrain the PDF's
of the proton, notably for the gluon at high proton momentum fractions. 
Therefore this analysis provides valuable input to determine the strong coupling at high 
momentum scales. The impact on the extraction of the PDF's is investigated.
The inclusion of CMS inclusive jet data in comparison to HERA inclusive DIS data alone
significantly reduces the uncertainty on the gluon distribution for fractional parton 
momenta $x \gtrsim 0.01$. In particular model and parameterisation uncertainties of the up and 
down quark distributions decrease for $x$ larger than 0.3. The effect is more pronounced for 
the up quark.

The jet data of this analysis has been included to fit the strong coupling constant as a 
free parameter in addition to the PDF's. This is not possible with HERA inclusive DIS data 
alone because of too large correlations between
$\alpha_S(M_Z)$ and the gluon PDF. The result is consistent with dedicated fits of $\alpha_S(M_Z)$
making use of global fit PDFs, where the gluon is much better constrained
through the data of various experiments and several observables. The PDF sets CT10~\cite{ct10}
and MSTW2008~\cite{mstw2008a}~\cite{mstw2008b} are used therefore at NLO evolution order. All 
fits with the MSTW2008-NLO, MSTW2008-NNLO, and NNPDF2.1-NNLO PDFs give compatible results for
both PDF sets. Cross checks with the CT10 PDFs at NNLO evolution order or with
additional corrections for parton shower effects are consistent within errors. 
Using these theory predictions the strong coupling constant is determined from the
inclusive jet cross section to be  
$\alpha_s(M_Z) = 0.1185 \pm 0.0019(\mbox{exp.}) \pm 0.0028(\mbox{PDF}) \pm 0.0004(\mbox{NP})^{+0.0055}_{-0.0022}(\mbox{scale})$, 
which is in agreement with the world average~\cite{pdg2012} $\alpha_s(M_Z)=0.1184 \pm 0.0007$.
To test the running of the strong coupling all fits have also been carried out for six bins in 
inclusive jet $p_T$ separately, where the scale $Q$ of $\alpha_S(Q)$
is identified with $p_T$. The observed behaviour of $\alpha_S(Q)$
is consistent with the momentum scale dependence predicted by the Renormalisation
Group Equation (RGE) of QCD and extend the H1, ZEUS, and D0 results to the 1 TeV range.

The cross sections of various top quark and boson production processes have been measured 
in $pp$ collisions at $\sqrt{s}=$ 7 and 8 TeV, with and without additional jets.
In Fig.~\ref{SMprocesses} (left) the CMS data are shown in good agreement with the 
theory prediction of the Standard Model. Limits for the processes $\gamma\gamma\rightarrow WW$
and $WV\gamma$ production, where $V$ is one of the vector bosons $W$, $Z$, are indicated, too.

The shape of the differential cross section for the transverse momentum distribution of 
dimuons produced in the Drell-Yan process has been measured~\cite{pas-smp-12-025} in the $Z$
boson mass range of 60 to 120 GeV in $pp$ collisions at $\sqrt{s} = 8$ TeV.
The results are obtained using a dedicated data sample with a reasonably low number of multiple
interactions per bunch crossing, corresponding to an integrated luminosity of  18.4 pb$^{-1}$.
The measured transverse momentum distributions are compared to common used event generators
like PYTHIA~\cite{pythia6} and several tunes including Z2star for the low transverse momentum 
($q_T$) region and to the theory predictions of FEWZ~\cite{fewz} at ${\cal{O}}(\alpha_S^2)$ 
for the high $q_T$ region. The low momentum 
region discriminates models of soft interaction including the showering scheme. 
Only moderate agreement between data and POWHEG~\cite{powheg} + PYTHIA Z2star for
$q_T$ below 20 GeV can be observed, whereas the description of PYTHIA with Z2star tune show
better agreement. On the high $q_T$ side of the spectrum better agreement of data with the 
predictions from the MADGRAPH~\cite{madgraph} generator are observed. However, in the range 
of $q_T$ beyond 100 GeV, the accuracy in data is not sufficient to significantly discriminate 
among available theoretical calculations of varying accuracy. In conclusion, the
PYTHIA generator with Z2star tune is able to describe data well in the low $q_T$ region, 
while MADGRAPH predictions are in general in good agreement with data at high $q_T$. 
The measurements done at 8 TeV are also compared with the ones at 7 TeV in terms
of the ratio of the cross sections. An overall good agreement is observed in the ratio of 
data and the predictions from the theory.

A measurement of inclusive W and Z boson production cross sections in pp collisions at
$\sqrt{s} = 8$ TeV is presented~\cite{pas-smp-12-011}. 
Electron and muon final states are analyzed in a data
sample collected with the CMS detector corresponding to an integrated luminosity of
$18.7\pm 0.9$ pb$^{-1}$.  The $W$ and $Z$ bosons are observed via their decays into 
electrons and muons. Their measured inclusive cross sections are
$\sigma(pp\rightarrow WX) \times Br(W\rightarrow\ell\nu) = 11.88 \pm 0.03(\mbox{stat})
\pm 0.22(\mbox{syst}) \pm 0.52(\mbox{lumi})$ nb and
$\sigma(pp\rightarrow ZX) \times Br(Z\rightarrow\ell^+\ell^-) = 1.12 \pm 0.01(\mbox{stat})
\pm 0.02(\mbox{syst}) \pm 0.05(\mbox{lumi})$ nb limited to a dilepton mass
range of 60 to 120 GeV. Ratios of cross sections are also produced.
The measurements are consistent between the electron and muon channels, and in agreement
with Next-to-Next-to-Leading Order (NNLO) QCD cross-section calculations.

A study of the $WV\gamma$ three electro-weak boson production and the anomalous quartic gauge 
boson couplings have been measured~\cite{pas-smp-13-009}, based on events containing a 
$W$ boson decaying to leptons, 
a second vector boson $V$ ($W$ or $Z$ boson) decaying to two jets, and a photon. 

Together with the triple $WW\gamma$ and $WWZ$ gauge boson vertices, the SM predicts the 
existence of quartic $WWWW$, $WWZZ$, $WWZ\gamma$ and $WW\gamma\gamma$
vertices. The direct investigation of gauge boson self-interactions provides a crucial
test on the gauge structure of the SM and is expected to play a significant role at LHC 
energies.
The data analyzed corresponds to 19.3 fb$^{-1}$ of integrated luminosity from pp collisions at
$\sqrt{s} = 8$ TeV collected in 2012 by the CMS detector at the Large Hadron Collider. 
For this amount of data an upper limit of 241 fb is set at 95\% confidence level for $WV\gamma$
production with photon $p_T > 10$ GeV. No evidence of anomalous $WW\gamma\gamma$ and $WWZ\gamma$
quartic gauge couplings is found. The following 95\% confidence level upper limits
are obtained for these couplings:
$-21 < a^W_0 / \Lambda^2 < 20$ TeV$^{-2}$,
$-34 < a^W_C / \Lambda^2 < 32$ TeV$^{-2}$,
$-25 < f_{T,0} / \Lambda^4 < 24$ TeV$^{-4}$,
$-12 < \kappa^W_0 / \Lambda^2 < 10$ TeV$^{-2}$ and
$-18 < \kappa^W_C / \Lambda^2 < 17$ TeV$^{-2}$.
A comparison to previous measurements of the LEP experiment L3 for the vertex $WW\gamma$
and of the D\O\ experiment for the vertex $\gamma\gamma\rightarrow WW$ is given in 
Fig.~\ref{SMprocesses} (right).
This is the first time ever that limits on $f_{T,0}$ and the CP conserving couplings
$\kappa^W_0$ and $\kappa^W_C$ are reported.

\begin{figure}[t]
 \vspace*{-4ex}
 \resizebox{0.63\columnwidth}{!}{%
 \hspace*{-4ex} \includegraphics{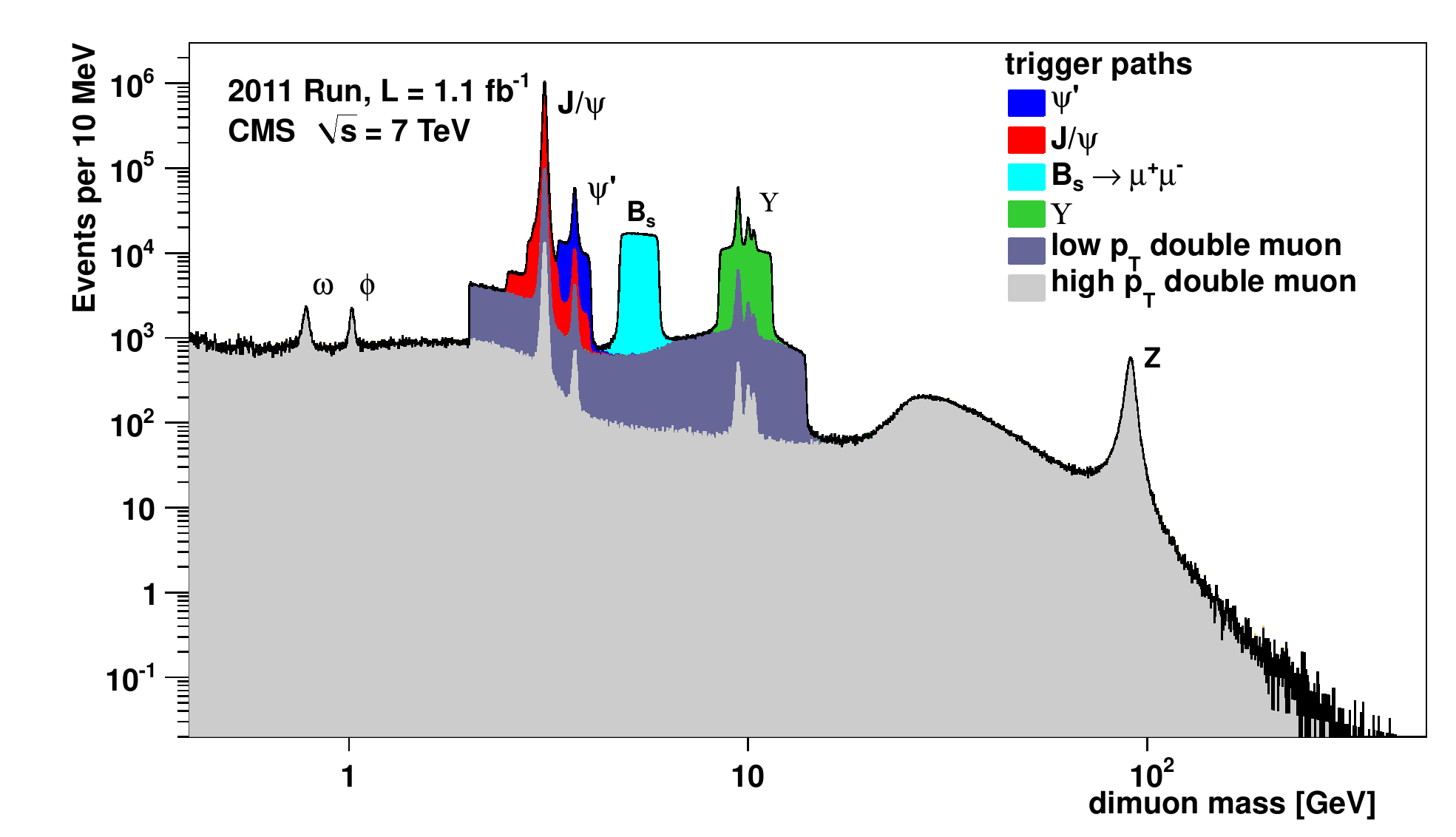} 
 \hspace*{-4ex}
}
 \unitlength 1cm
 \begin{picture}(10.,6.)
\put(0.2, 5.77){\includegraphics[bb=110 40 496 413, width=5.6cm, angle=270, clip]{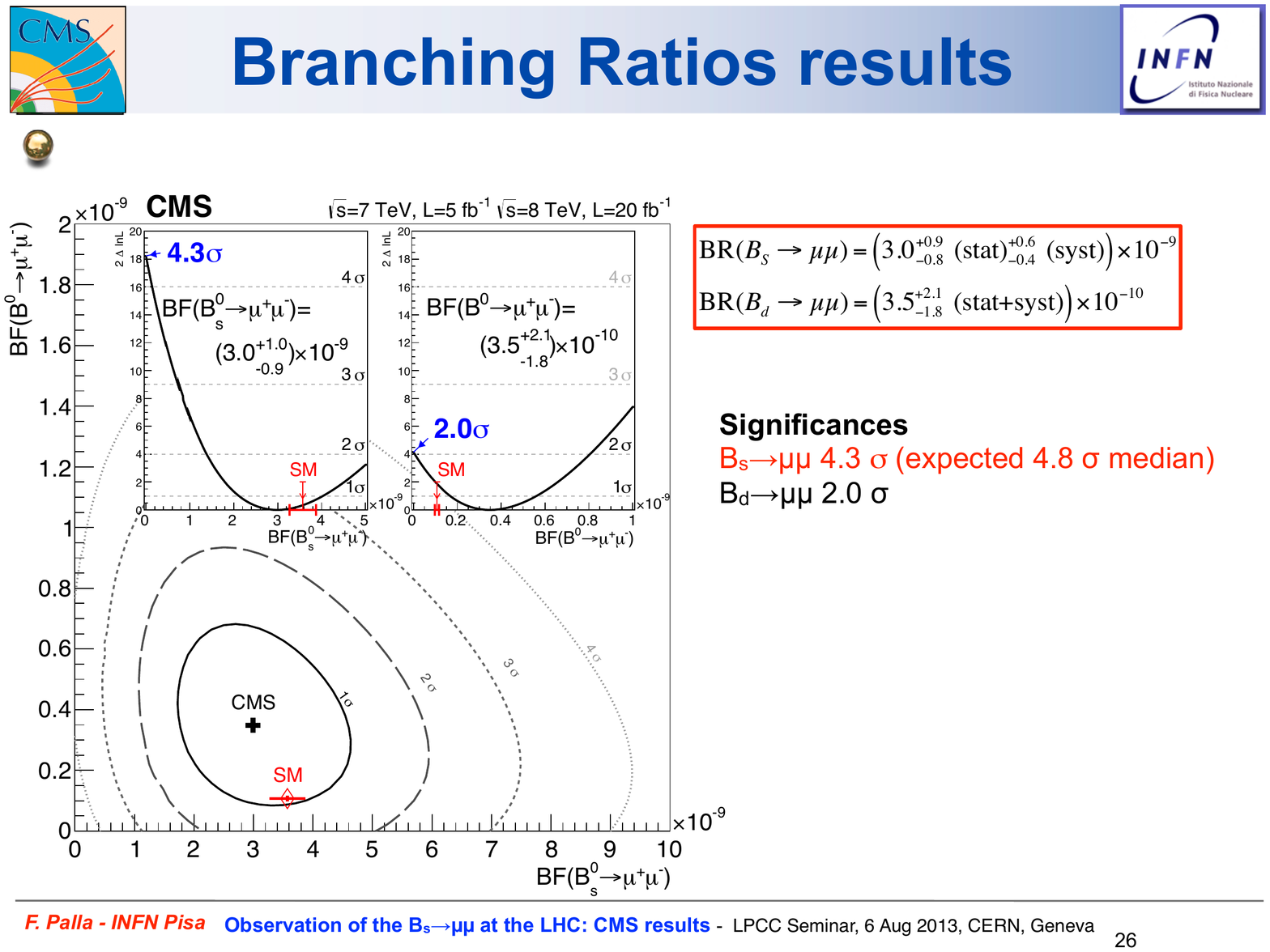} }
\put(5.62, 0.7){\scalebox{0.5}{$10^{-9}$}}
\end{picture}

\caption{ \label{BsPhysics}
Left: Dimuon invariant mass spectrum. Shown ar dedicated low $p_T$ threshold dimuon triggers with a narrow mass window around resonances like $B_s$. Right: CMS obtained values for the branching fractions
of $B_d\rightarrow\mu^+\mu^-$ and $B_s\rightarrow\mu^+\mu^-$. The insets show the likelihoods of 
the one dimensional projections with the corresponding significances. The SM expectation value is 
indicated in red.}
\end{figure}

\section{Heavy flavour physics} \label{HFphysics}
Dedicated low $p_T$ threshold dimuon triggers with a narrow mass window around resonances
such as $J/\psi$, $\psi'$, $B_s$ and $\Upsilon$ have been exploited in low luminosity
periods with relatively low number of pile-up events per proton bunch crossing.
The invariant dimuon mass spectrum for the different applied dimuon triggers is shown in 
Fig.~\ref{BsPhysics} (left). Of particular interest here is the measurement of the
Branching Fraction (BF) of the decay $B_s\rightarrow\mu^+\mu^-$ which is highly suppressed in 
the SM and new particles could contribute in loop and box diagrams and enhance the BF.
Within the Minimal Supersymmetric SM (MSSM) the BF is proportional to the sixth power of the
ratio $\tan\beta$ of the two Higgs doublets vacuum expectation values. 
The likelihood for the BF's obtained by CMS for the two processes $B_d\rightarrow\mu^+\mu^-$ and 
$B_s\rightarrow\mu^+\mu^-$ exploiting the full integrated luminosity of 5 fb$^{-1}$ at
$\sqrt{s} = 7$ TeV and 20 fb$^{-1}$ at $\sqrt{s} = 8$ in $pp$ collisions are shown in 
Fig.~\ref{BsPhysics} (right) in agreement with the SM expectation. 
The numerical results are $BF(B^0_s\rightarrow \mu^+\mu^-) = 2.9^{+1.1}_{-1.0}(\mbox{stat})^{+0.3}_{-0.1}(\mbox{syst})\times 10^{-9}$, corresponding to a $4.0$ standard deviations evidence and
$BF(B^0_d\rightarrow \mu^+\mu^-) = 3.5^{+2.1}_{-1.8}\times 10^{-10}$, corresponding to a $2.0$ standard deviations significance. 
In assuming a Null hypothesis the later BF measurement can also be expressed
in terms of a limit, namely $BF(B^0_d\rightarrow \mu^+\mu^-) < 7.4 \times 10^{-10}$ 
at 95\% Confidence Level (C.L.).
In combination with the LHCb results exploiting an integrated luminosity of 3 fb$^{-1}$,
the decay $B_s\rightarrow\mu^+\mu^-$ has been observed with a Branching Fraction of
$BF(B^0_s\rightarrow \mu^+\mu^-) = (2.9\pm 0.7) \times 10^{-9}$ in consistence with the SM
expectation and the combination for the decay $B_d\rightarrow\mu^+\mu^-$ yields
$BF(B^0_d\rightarrow \mu^+\mu^-) = 3.6^{+1.6}_{-1.4}\times 10^{-10}$ which is not yet statistically
significant but again consistent with the SM expectation.

\section{Top quark physics}
The top quark is the heaviest known particle to-date and it is produced 
pair-wise or singly with associated particles.
The $t\bar{t}$ production at the LHC in $pp$ collisions is dominated by gluon 
gluon fusion (~90\%), the remainder being due to quark antiquark annihilation 
(~10\%). At a centre-of-mass energy of $\sqrt{s}=7$~TeV the NLO cross 
section~\cite{mc@nlo1} yields 158~pb and the approximated NNLO cross 
section~\cite{ttbarNNLO} 163~pb. At $\sqrt{s} = 8$ TeV the approximated NNLO cross 
section amounts to 246 pb.
In the SM the top quark decays almost exclusively into a $W$ boson and a $b$
quark. The $W$ boson in turn decays in 2/3 of cases into a $q\bar{q}$ pair
and in 1/3 of cases leptonically. Single top quark production is dominated by
the $t$-channel with a cross section of 64~pb at $\sqrt{s}=7$~TeV and 87~pb at
$\sqrt{s}=7$~TeV, followed by the $tW$-channel with 15.7~pb (22.4~pb) at $\sqrt{s}=7(8)$ TeV
and the $s$-channel with 4.6~pb (5.6~pb) at $\sqrt{s}=7(8)$ TeV. 

\begin{figure}[b]
\vspace*{-2.0ex}
\resizebox{0.5\columnwidth}{!}{%
 \hspace*{-6.0ex} \includegraphics{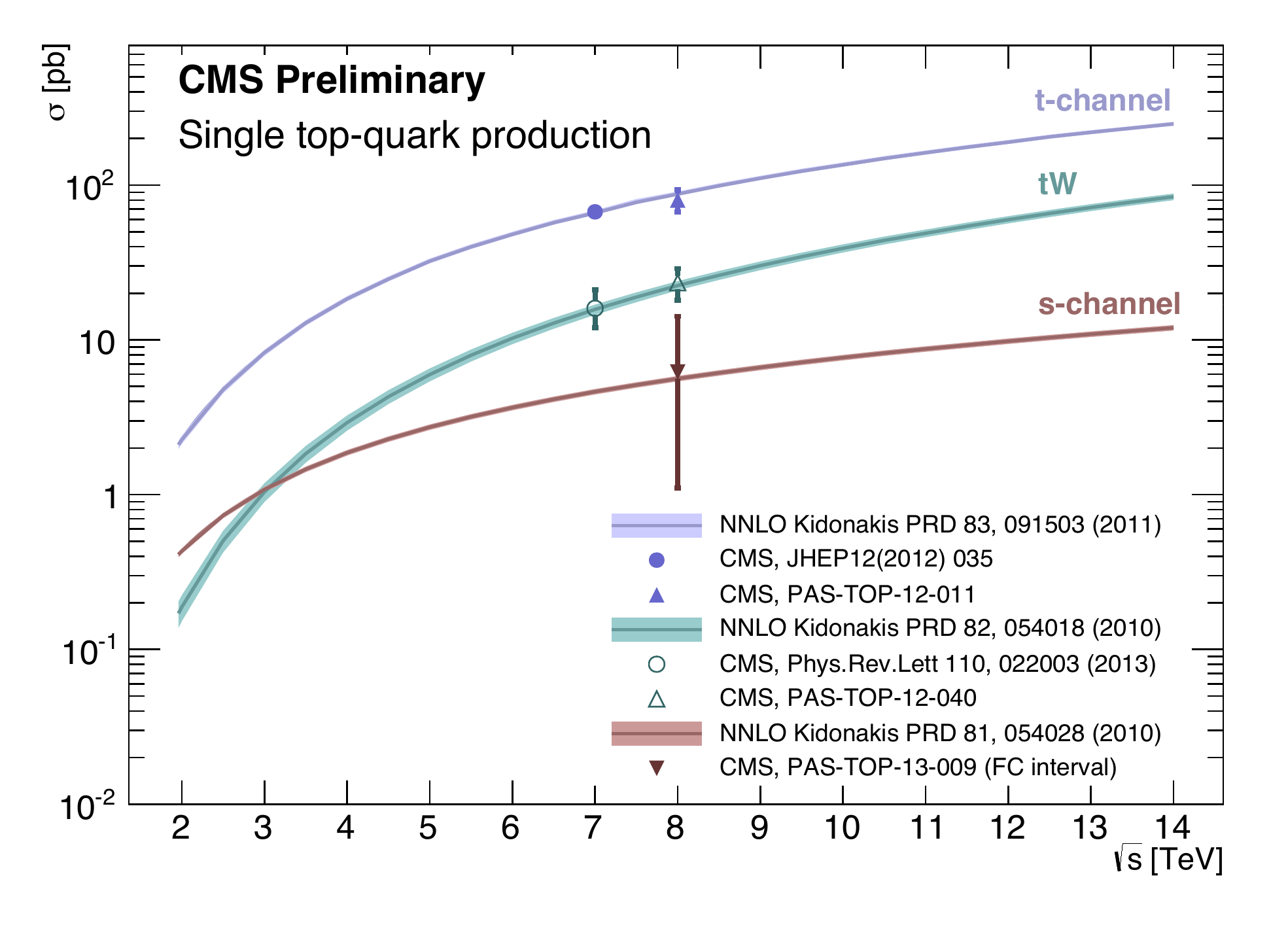} 
 }
 \hspace*{-40ex} 
 \unitlength 1cm
 \begin{picture}(10., 3.)
 \put(6.2, -0.1){ \includegraphics[width=8.5cm]{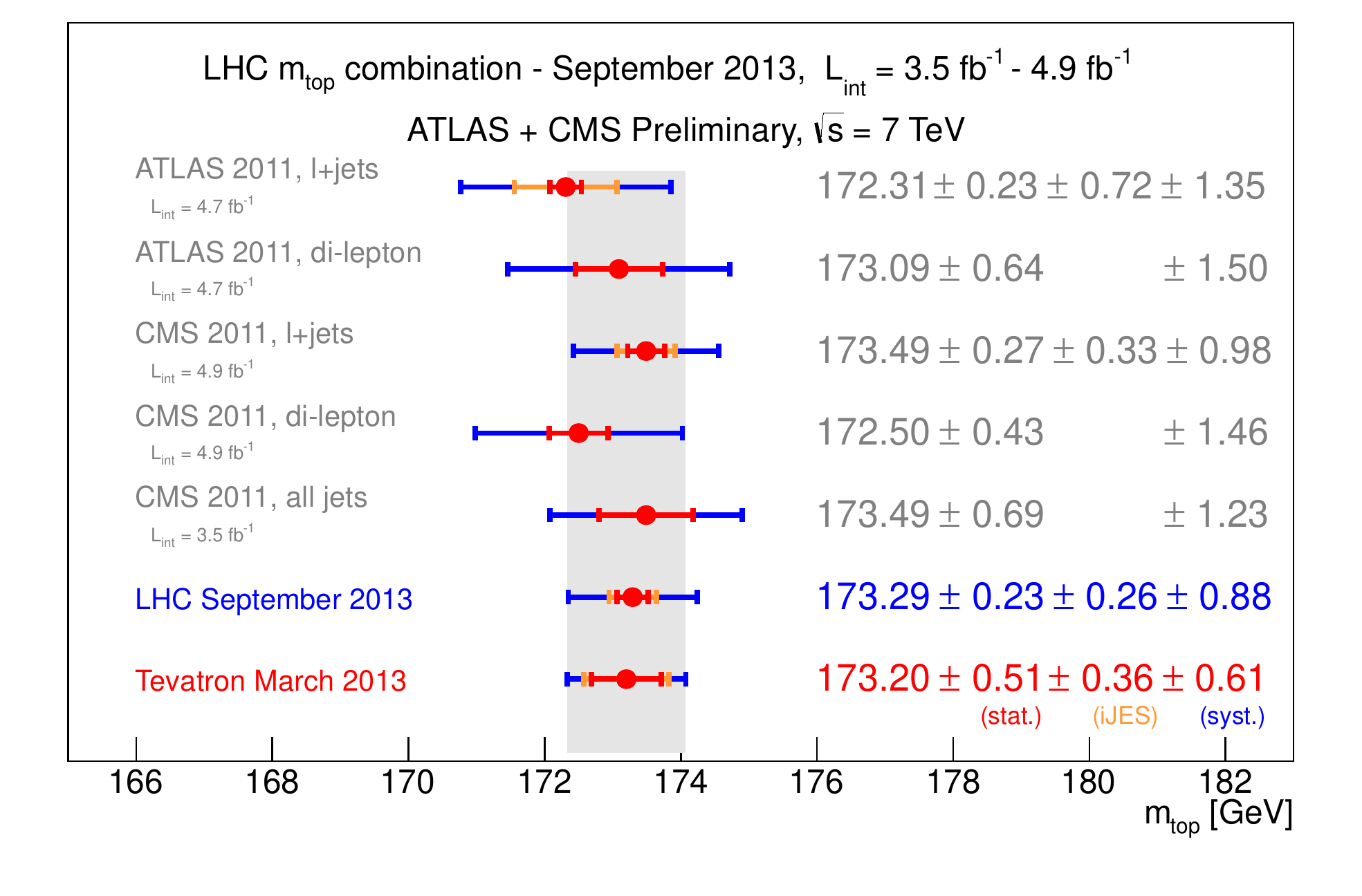} }
 \end{picture}

\caption{ \label{top_summary}
Left: Measured cross sections of single top production for the different production channels
at $\sqrt{s} = 7$ and 8 TeV in comparison to the NNLO predictions. 
Right: LHC top mass combination in comparison to the Tevatron combination.
}
\end{figure}

The $t\bar{t}$ cross section measurements are done in all 
possible decay channels by means of a binned likelihood fit at $\sqrt{s} = 7$ and 8 TeV.
The $e/\mu$ + jets channel~\cite{pas-top-12-006} yields a cross section of
$\sigma(t\bar{t})= 228 \pm 9(\mbox{stat}) ^{+29}_{-26}(\mbox{syst}) \pm 10(\mbox{lumi})$ pb,
the $ee$, $e\mu$, $\mu\mu$ channel~\cite{pas-top-12-007} yields 
$\sigma(t\bar{t})= 227 \pm 3(\mbox{stat}) \pm 11(\mbox{syst}) \pm 10(\mbox{lumi})$ pb
which at the same time corresponds to the numbers of the combination of the two channels
due to the larger errors of the former one and the agreement between the channels which
amounts to the 5 - 15\% level of experimental uncertainty.
The measurements are compatible with the theory prediction in approximated NNLO with scale and 
PDF uncertainties of the order 10\%. A full NNLO prediction is available and reduces this
uncertainty to about 4\%. Therefore the cross section measurements can be exploited to probe
the gluon density of the proton. 
Single top quark production has been measured
in the $t$-channel with the $W$ boson decaying leptonically 
at $\sqrt{s} = 7$~\cite{CMS_JHEP12_2012_035} 
yielding $\sigma=70.2\pm 5.2(\mbox{stat})\pm 10.4(\mbox{syst})\pm 3.4(\mbox{lumi})$~pb
and at $\sqrt{s} = 8$ TeV~\cite{pas-top-12-011}
yielding $\sigma = 80.1 \pm 5.7(\mbox{stat}) \pm 11.0(\mbox{syst}) \pm 4.0(\mbox{lumi})$~pb.
The $t$-channel single top production is also being exploited
for a direct measurement of the CKM matrix element 
$|V_{tb}|=\sqrt{\sigma_{t}/\sigma^{\mbox{\scriptsize theo}}_{t}(|V_{tb}\equiv 1|)} = 1.04 \pm 0.09(\mbox{exp})\pm 0.02(\mbox{theo})$.
In the $tW$-channel the cross section has been observed~\cite{pas-top-12-040} at 
$\sqrt{s} = 8$ TeV to be $\sigma = 23.4 ^{+5.5}_{-5.4}$~pb with a significance of 6.0$\sigma$ 
and in agreement with the SM prediction.
At $\sqrt{s} = 7$ TeV a $4.0$ standard deviations evidence~\cite{pas-top-11-022} for $tW$ production has 
been obtained with a cross section of $\sigma = 16^{+5}_{-4}$~pb in agreement with the SM 
prediction. The 2012 dataset at $\sqrt{s} = 8$ TeV with an integrated luminosity of 19.3 fb$^{-1}$
leads to an upper limit on the $s$-channel cross section times branching ratio of 
11.5 pb at 95\% C. L.~\cite{pas-top-13-009}. 
Figure~\ref{top_summary}, left shows the single top cross 
section measurements of the different channels and centre-of-mass energies 
in comparison to the SM predictions.  
The right plot of Fig.~\ref{top_summary} shows the top mass measurements
in the different decay channels from CMS and ATLAS in comparison to theory.
The dilepton channel is treated as a counting experiment. The hadronic channel 
making use of an unbinned likelihood is taken as one bin.
The Best Linear Unbiased Estimate (BLUE) method~\cite{blue} is used as a cross 
check for the combination of the different channels taking into account 
correlations between different contributions to the measurements.
The top mass combination of CMS and ATLAS for the 2011 dataset at $\sqrt{s} = 7$ TeV
yields
$m_{\mbox{\scriptsize top}}=173.29 \pm 0.23(\mbox{stat})\pm 0.26(\mbox{iJES}) \pm 0.88(\mbox{syst})$~GeV~\cite{pas-top-13-005} where iJES indicates the error due to intercalibration Jet Energy Resolution due to modelling of radiation in the jet differences in $\eta$ and $p_T$. 
This result is competitive with the Tevatron combination
$m_{\mbox{\scriptsize top}}=173.20 \pm 0.51(\mbox{stat})\pm 0.36(\mbox{iJES}) \pm 0.61(\mbox{syst})$~GeV. The systematic error is below 1 GeV and the statistical and iJES uncertainties are smaller 
than the Tevatron combination. 

\begin{table}[b]
 \unitlength 1cm
 \begin{picture}(10., 6.2)
 \put(-0.02, -0.2){ \includegraphics[bb = 31 413 372 567, width=15.0cm, clip]{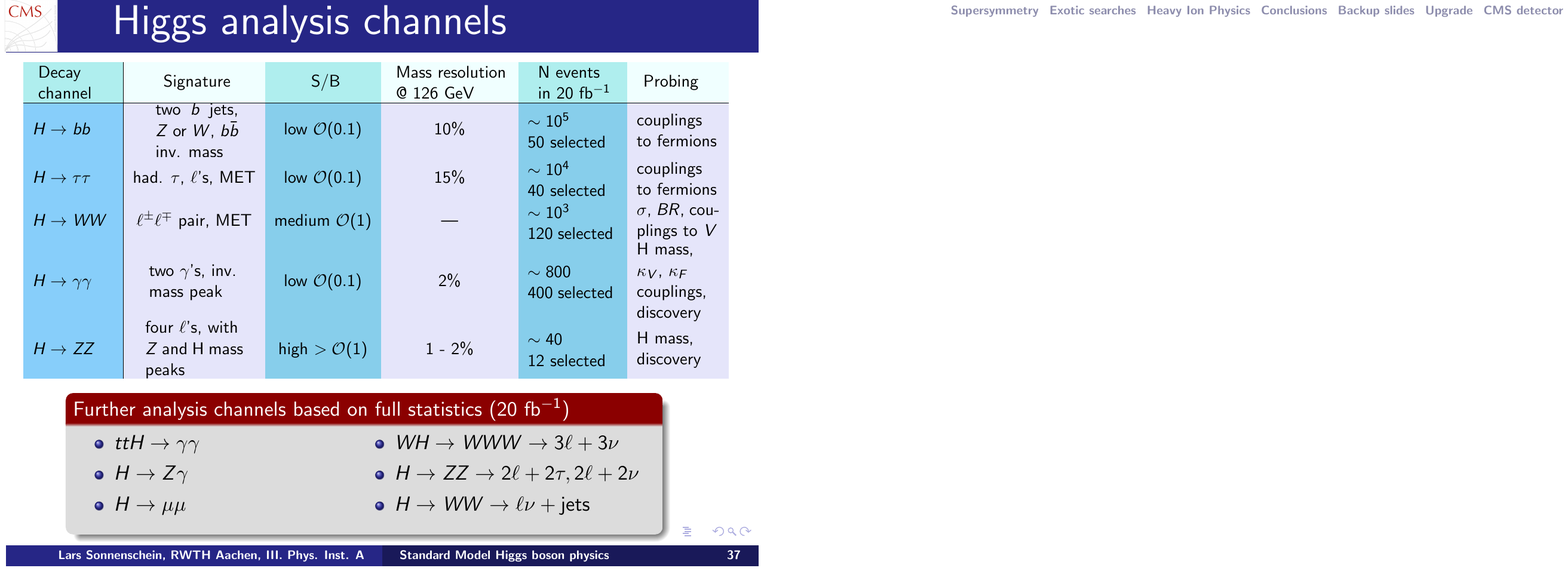} }
 \end{picture}

\caption{ \label{higgs_channels}
Characteristics of the most important Higgs channels analysed by the CMS experiment.
}
\end{table}

\begin{figure}[b]
\vspace*{-2.5ex}
\resizebox{0.58\columnwidth}{!}{%
 \hspace*{-4ex} \includegraphics{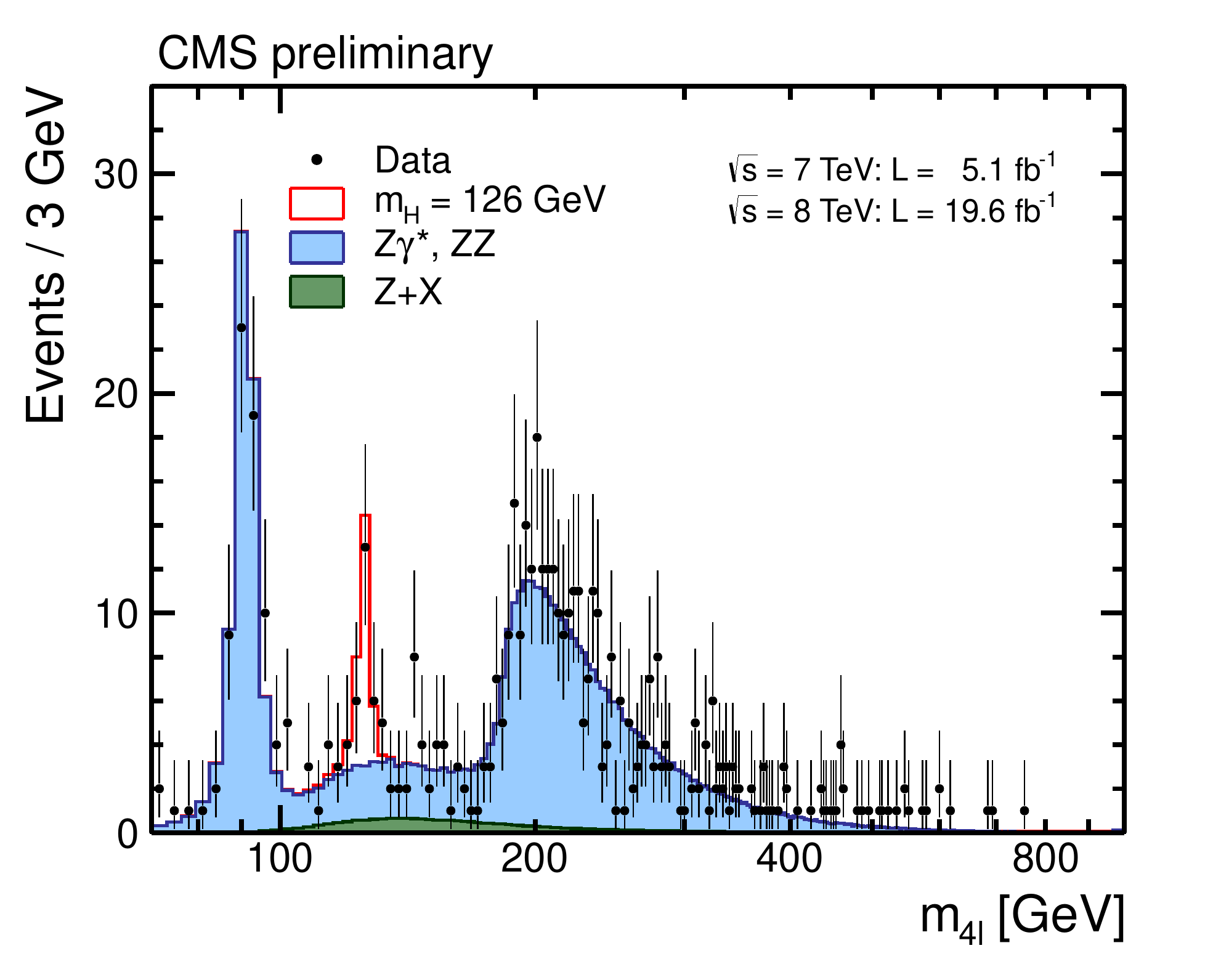} 
 }
 \hspace*{-40ex} 
 \unitlength 1cm
 \begin{picture}(10., 3.)
 \put(6.2, 0.11){ \includegraphics[width=7.19cm]{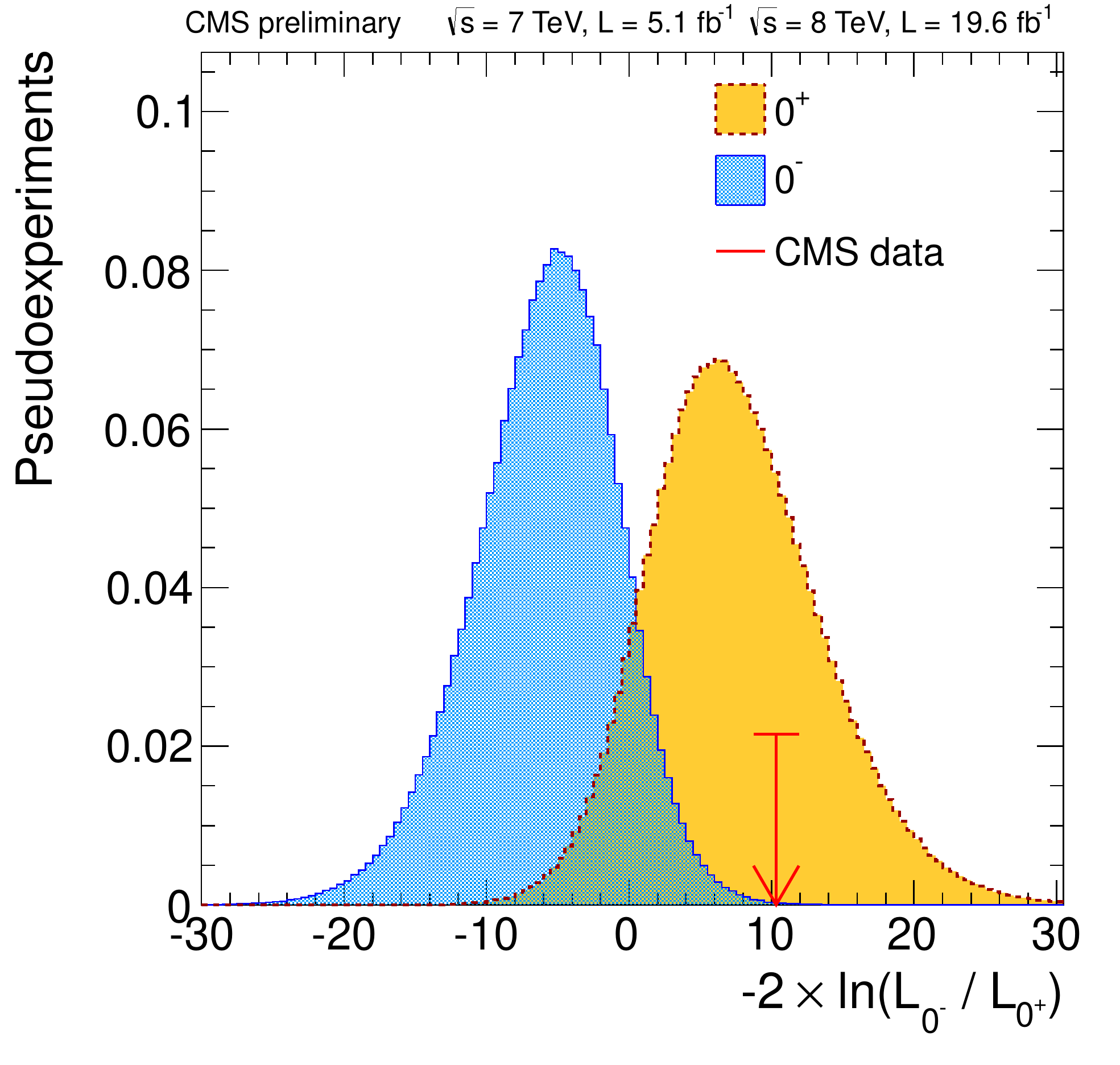} }
 \end{picture}

\vspace*{-2ex} 
\caption{ \label{higgs_ZZto4l}
Left: Invariant mass distribution of the four leptons indicating clearly the peak at 
$m_H=125.8$~GeV, exploiting the statistics of the full 2011 (5.1 fb$^{-1}$ at $\sqrt{s} = 7$ TeV) 
and 2012 (19.6 fb$^{-1}$ at $\sqrt{s} = 8$ TeV) datasets. 
Right: Measurement of the spin-parity of the found Higgs boson, strongly favouring a scalar
Higgs boson and disfavouring a pseudoscalar Higgs boson.
}
\end{figure}

\section{Standard Model Higgs boson physics}

The electro-weak symmetry breaking mechanism, also referred to as 
Brout-Englert-Higgs (BEH) mechanism has been primarily conceived to endow the 
particles with mass. Experimental analysis of this mechanism will allow 
to probe the mathematical consistency of the Standard Model (SM) of particle physics 
beyond the TeV energy scale where divergences of the theoretical description can be 
prevented by means of loop corrections taking contributions and couplings to
a scalar Higgs boson into account. 
Several production mechanisms contribute at the LHC, namely the dominant gluon fusion
process, the subdominant vector boson fusion and the associated production process.
The associated production via a vector boson suffers a high level of multijet 
background and is less favoured compared to the Tevatron where the reach
is limited to relatively low Higgs boson masses~\cite{Sonnenschein}.
The associated production in $t\bar{t}$ events is too small for reach
within the existing datasets. Present CMS exclusion limits for $t\bar{t}H$ production
with $H\rightarrow\gamma\gamma$ are about five times above the SM 
expectation~\cite{pas-hig-13-015}.

The Higgs boson Yukawa coupling strength is proportional to the mass of the 
particles, favouring the coupling via a top quark loop.
Photons and gluons can only couple indirectly.
The decay modes and branching ratios depend on the Higgs boson mass. 
At the mass of 126 GeV $H\rightarrow b\bar{b}$ is the dominant decay channel,
followed by the decays into $WW$, $gg$, $\tau\tau$, $c\bar{c}$, $ZZ$, $\gamma\gamma$ 
and $Z\gamma$.
Most decay modes are pursued by CMS. Table~\ref{higgs_channels} shows for the most 
important Higgs decay channels analysed by CMS the signal signatures, 
the Signal over Background (S/B) ratio,
the mass resolution at $m_H = 126$ GeV, the number of events with an integrated luminosity
of 20 fb$^{-1}$ and Higgs boson properties which can be probed. 

The CMS data has been analysed to study a standard model Higgs boson in the four-lepton
decay channel $H\rightarrow ZZ \rightarrow 4\ell$ ($\ell=e,\mu$)
and $H \rightarrow ZZ \rightarrow 2\ell\tau$~\cite{pas-hig-13-002}. 
The mass distributions are measured with
four-lepton invariant masses above a threshold of 100 GeV exploiting 5.1 fb$^{-1}$ of
integrated luminosity at $\sqrt{s} = 7$ TeV and 19.6 fb$^{-1}$ at $\sqrt{s} = 8$ TeV.
To discriminate background, information on an event-by-event basis is used from the measured 
four-lepton mass, the mass uncertainty, a kinematic discriminant and information sensitive to
the production mechanism, such as associated dijet characteristics and transverse momentum
of the four-lepton system. Upper limits at 95\% C.L. exclude the SM-like Higgs boson in the 
range of $130 - 827$ GeV while the expected exclusion range amounts to  $113.5 - 778$ GeV. 
The new boson discovered by the CMS and ATLAS experiments is observed in the 4 lepton
channel measured by CMS, 
with a local significance of 6.7 standard deviations above the expected background. 
Figure~\ref{higgs_ZZto4l} (left) shows the invariant four lepton mass distribution 
of signal compared to the expected SM background.
A measurement of its mass yields $125.8 \pm 0.5(\mbox{stat}) \pm 0.2(\mbox{syst})$ GeV. 
The signal strength $\mu$ relative to the expectation for a standard model Higgs boson, 
is measured to be $\mu = 0.91^{+0.30}_{-0.24}$ at the mass where the boson has been found. 
The signal strength modifiers associated with vector bosons and fermions in the production
mechanism are measured to be $\mu_V = 1.0^{+2.4}_{-2.3}$ and $\mu_F = 0.9^{+0.5}_{-0.4}$ 
in consistence with the SM expectations.
The spin-parity of the boson is studied and the pure scalar hypothesis is found to be 
consistent with the observation. The measured state has been compared to six other 
spin-parity hypotheses including spin 0, 1 and 2 states with positive and negative spin-parity. 
Figure ~\ref{higgs_ZZto4l} (right) shows the comparison of the scalar and pseudoscalar
Higgs boson hypotheses. The CMS measurement is indicated by the red arrow and clearly
disfavours a pseudoscalar Higgs boson.
The fraction of a CP-violating contribution to the decay amplitude, expressed through the 
fraction $f_{a3}$ of the corresponding decay rate, is measured to be $f_{a3} = 0.00^{+0.23}_{-0.00}$
again consistent with the SM expectation. The data disfavour the pure pseudoscalar hypothesis 
0 with a $CL_s$ value of 0.16\%. This disfavours the pure spin-2 hypothesis of a narrow 
resonance with minimal couplings to the vector bosons with a $CL_s$ value of 1.5\%. 
The spin-1 hypotheses are disfavoured with an even higher level of confidence.

Evidence for the standard model Higgs boson decaying into a pair of $\tau$ leptons has been 
established~\cite{pas-hig-13-004}. The analysis is based on the full $pp$ collision data sample 
recorded by CMS in 2011 and 2012, corresponding to an integrated luminosity of 4.9 fb$^{-1}$
at a $\sqrt{s} = 7$ TeV and 19.4 fb$^{-1}$ at $\sqrt{s} = 8$ TeV. It is performed in five 
different final states, $\mu\tau_h$, $e\tau_h$, $e\mu$, $\tau_h\tau_h$ and $\mu\mu$.
These channels are combined with the result of a search for the standard model Higgs boson
decaying into a pair of $\tau$ leptons where the Higgs boson is produced in association with 
$W$ or $Z$ bosons decaying leptonically. 
An excess of events is observed over the expected background contributions,
with a maximum local significance of $3.2$ standard deviations 
for $m_H$ values between 115 and 130 GeV.
The best fit of the observed $H\rightarrow\tau\tau$ signal cross section for
$m_H = 125$ GeV is $0.78 \pm 0.27$ times the SM expectation. These
observations constitute evidence for the 125 GeV Higgs boson decaying to a pair of
$\tau$ leptons.

The Higgs signal strengths $\mu = \sigma/\sigma_{SM}$ have been measured in many decay channels
and associated production modes. The combination of the best fit values amounts to
$\mu = 0.80 \pm 0.14$ for a Higgs boson mass of $m_H = 125.7$ GeV.
The couplings strengths to fermions $\kappa_F$ and to bosons $\kappa_V$ have also been measured
and the combination of various analysis channels are close to the SM expectation.
Systematic and theoretical uncertainties arising e.g. from QCD scale, PDF's and $Br$'s
are taken into account.

Projections of Higgs boson measurements with 300 fb$^{-1}$ at $\sqrt{s} = 14$ 
TeV~\cite{cms-note-2012-006} have been studied. 
The performance estimation is based on extrapolation of ICHEP 2012 performance.
There is potential to shrink the errors by a factor 2 - 4 on SM Higgs boson signal
and coupling strengths.

\begin{figure}[t]
 \unitlength 1cm
 \begin{picture}(10.0, 11.9)
 \put(-0.45, -0.1){ \includegraphics[width=8.4cm]{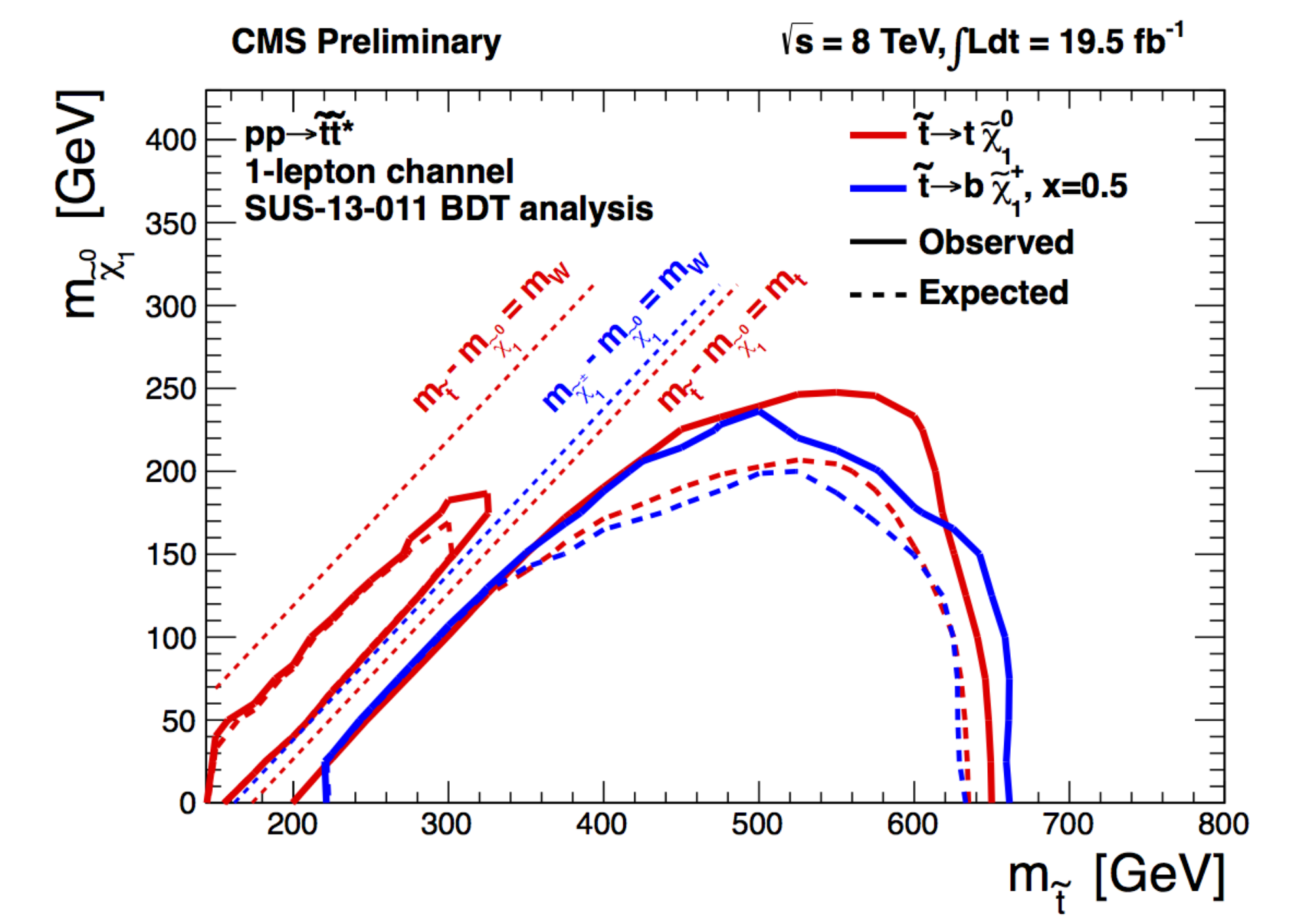} }
 \put(7.9, -0.2){ \includegraphics[width=7.8cm]{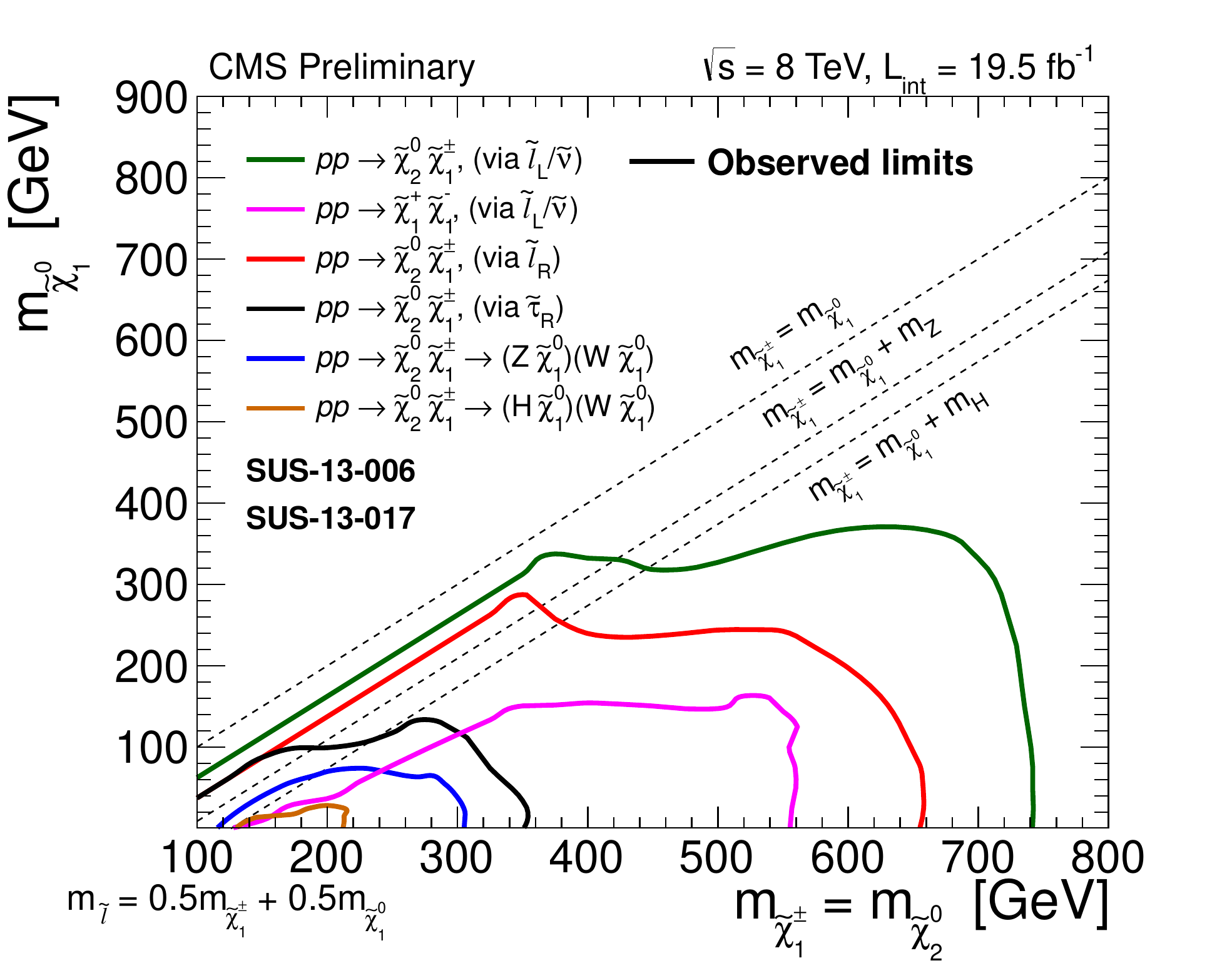} }
 \put(0.0, 6.1){ \includegraphics[width=6.45cm]{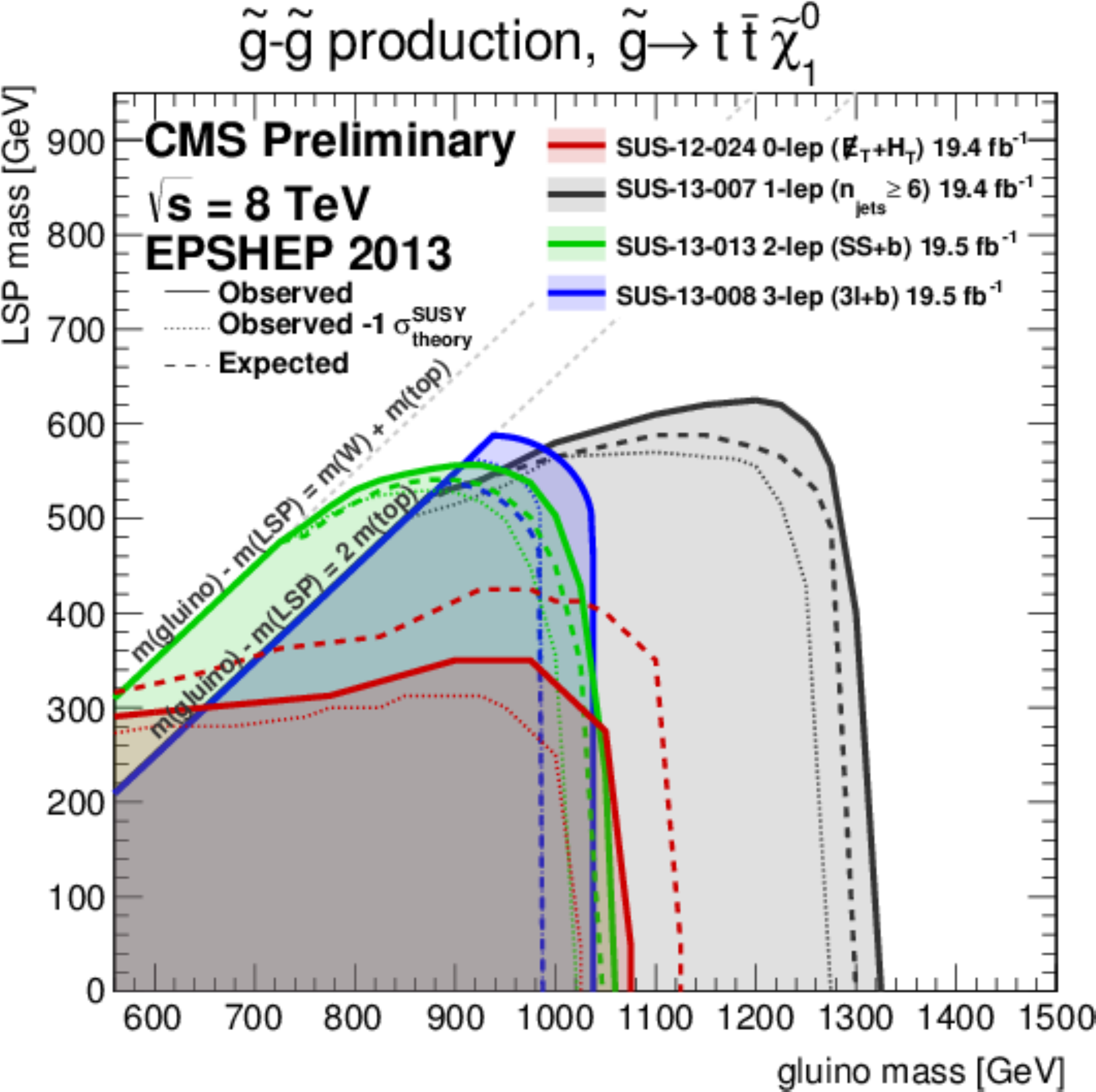} }
 \put(6.7, 6.1){ \includegraphics[width=8.9cm]{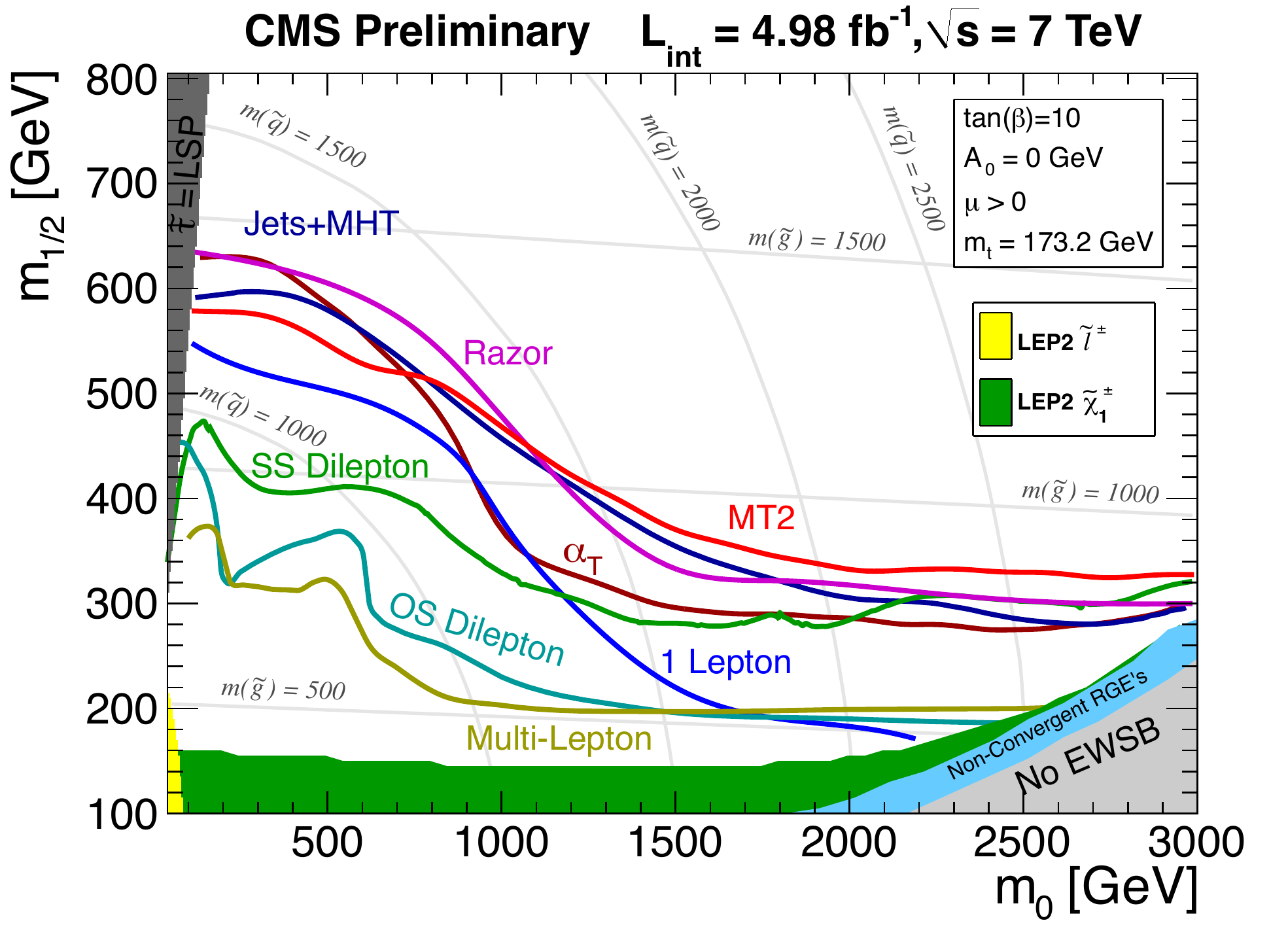} }
 \end{picture}

\caption{ \label{susy_summaryPlots}
SUSY exclusion summaries. Top left: Limits on gluino pair production 
($\tilde{g}\rightarrow t\bar{t}\tilde{\chi}^0_1$). 
Top right: CMSSM mSUGRA limits in the plane of gaugino mass $m_{1/2}$ vs. scalar mass $m_0$ for
$\tan\beta=10$, $A_0=0$ GeV, $\mu>0$ of various analyses on the 2011 dataset with 
$\sqrt{s} = 7$ TeV. Bottom left: Limits on direct stop production. 
Bottom right: Limits on electro-weak gaugino production.
}
\end{figure}

\section{Search for supersymmetry}
In the supersymmetric extension of the Standard Model corresponds
to each SM particle a supersymmetric partner particle with identical
gauge quantum numbers but spin differing by 1/2.
The primary motivation to introduce SUperSYmmetry (SUSY) is to solve the hierarchy problem.
Large fine-tuning is needed for physical parameters of a theory at high energy scale
in order to correctly describe the low-energy scale and light objects.
To keep the renormalised Higgs boson mass at the weak scale the bare mass requires an unnatural
fine-tuning of the order $(M_W/M_{GUT})^2\sim 10^{-28}$. 
Quadratic divergences due to loop corrections can be canceled by adding supersymmetric
partner particles. 
The highest expected production cross sections of supersymmetric particles at the 
LHC are given by gluino and squark pair production, followed by the electro-weak 
production of charginos and neutralinos.
The decay topology and kinematics is characterised by high transverse momentum jets,
$\not\!\!E_T$ due to the undetected Lightest Supersymmetric Particle (LSP) and
neutralino and chargino decays into SM fermion pairs and the LSP which serves as a
dark mater candidate in the case of $R$-parity conservation, where $R$-parity is defined
as $R_p= (-1)^{3(B-L)+2s}$ with $B$ = Baryon number, $L$ = Lepton number and $s$ = spin.
For supersymmetric particles holds $R_p=-1$ and for SM particles $R_p=+1$. 

In section~\ref{HFphysics} the observation of $B_S\rightarrow\mu\mu$ and its implications
for new physics, in particular for SUSY has been discussed. The measurement of the
branching ratio $Br(B_S\rightarrow\mu\mu)$ is consistent with the SM expectation 
which causes severe constraints on new physics. In particular for the
Constrained Minimal Supersymmetric SM (CMSSM) direct and indirect constraints suggest particle 
masses at the multi-TeV level. This implies that sparticle masses are mostly out of reach 
for the LHC. More severe, fine-tuning will be required for the CMSSM. 
Models beyond the CMSSM provide viable options. Notably third generation sfermions might be 
in reach. First and second generation sfermions can be made much heavier naturally.
Also the electro-weak SUSY sector might be in reach (light higgsino). Considerable phase space 
is accessible at the LHC.

A natural SUSY scenario with gauge-mediated SUSY breaking (GMSB) has been searched 
for~\cite{pas-sus-13-014} in the context of simplified model sparticle mass spectra (SMS),
where the potential signal is given by Feynman graphs describing the production mechanism and 
the decay chains. The sparticle masses are free parameters of such SMS.
The natural GMSB higgsino model chosen assumes that only the stop and the higgsino are 
accessible. The other sparticles are assumed to be located at higher mass scales out of reach
for the LHC. In the strong production a pair of stop quarks is produced with $b$ quarks in the
decay and Higgs bosons as well as the gravitino as LSP in the final state.
The branching ratio of the lightest neutralino into Higgs and gravitino is assumed to be one:
$Br(\tilde{\chi}^0_1\rightarrow H\tilde{G})=1$.
In the electro-weak production a pair of gauginos (charginos or neutralinos) is produced,
again with Higgs bosons and gravitino LSP's in the final state.
Events are selected in which there are two photons comprising a Higgs boson candidate, at
least two $b$-jets and missing transverse energy ($E_T^{\mbox{\scriptsize miss}}$). 
Exploiting an integrated luminosity of 19.5 fb$^{-1}$ in $pp$ collisions at $\sqrt{s} = 8$ TeV
no evidence of a signal has been found and lower limits at the 95\% C.L. on the stop mass 
beteen 360 to 410 GeV have been set, depending on the higgsino mass.

Searches for direct gluino pair production are accomplished in the 0-lepton 
channel~\cite{pas-sus-12-024} based on missing transverse energy $E_T^{\mbox{\scriptsize miss}}$
and transverse hadronic scalar momentum $H_T$, in the 1-lepton channel~\cite{pas-sus-13-007}
based on at least six jets in the hadronic final state, in the 2-lepton 
channel~\cite{pas-sus-13-013} based on Same Sign (SS) leptons and $b$-tagged jets
and in the 3-lepton channel~\cite{pas-sus-13-008} based on three leptons and $b$ tagged jets.
Limits are derived in the kinematic phase space of LSP mass vs. gluino mass $m_{\tilde{g}}$,
excluding gluino masses up to 1350 GeV.

Due to the strong constraints on the CMSSM from direct LHC searches and the Higgs boson
found at a mass of about 126 GeV it will not be able to solve the hierarchy problem anymore.
Therefore CMS does no longer provide CMSSM interpretations for searches accomplished from 
the $\sqrt{s} = 8$ TeV 2012 dataset on. Therefore the last CMS CMSSM mSUGRA limit interpretations
have been applied to the $\sqrt{s} = 7$ TeV 2011 dataset. The limits in the gaugino mass 
$m_{1/2}$ vs. scalar mass $m_0$ plane for $\tan\beta=10$, $A_0=0$ GeV, $\mu>0$ are summarized
in Fig.~\ref{susy_summaryPlots}, top right plot for various analyses
with different production mechanisms and final states.

A search for the direct pair production of top squarks, in the final state
consisting of a single isolated lepton, jets, large missing transverse energy
 and large transverse mass has been performed. 
Signal regions are defined by means of the output of a Boosted Decision Tree (BDT) 
multivariate discriminator and requirements on several kinematic discriminants. 
Two kinematic signal regions between the boundary conditions $m_{\tilde{t}}-m_{\tilde{\chi}^0_1} = m_W$
and $m_{\tilde{\chi}^{\pm}_1} - m_{\tilde{\chi}^0_1} = m_W$ on the one hand and beyond
$m_{\tilde{t}} - m_{\tilde{\chi}^0_1} = m_t$ on the other hand are considered as can be seen in 
Fig.~\ref{susy_summaryPlots}, top left plot.
The observed yields in the signal regions agree with the predicted backgrounds within 
uncertainties. The results are interpreted in the context of models of top-squark pair 
production and decay, probing top squarks with masses up to about 650 GeV.

Searches for the electro-weak production of gauginos (neutralinos $\tilde{\chi}^0$ and charginos
$\tilde{\chi}^{\pm}$) are accomplished with sleptons in the decay chain~\cite{pas-sus-13-006} 
giving rise to leptons and $E_T^{\mbox{\scriptsize miss}}$ in the final state. Another analysis is 
looking for a $W$ and a Higgs boson in the decay chain~\cite{pas-sus-13-017}, the $W$ boson
decaying leptonically and the Higgs boson giving rise to two $b$-jets.
LSP $\tilde{\chi}^0_1$ masses are excluded up to 350 GeV and chargino masses $\tilde{\chi}^{\pm}_1$
up to 740 GeV.
All approved CMS SUSY results and summaries are publically available at \newline
{\tt https://twiki.cern.ch/twiki/bin/view/CMSPublic/PhysicsResultsSUS}.

\begin{figure}[t]
 \vspace*{-3ex}
 \resizebox{0.5\columnwidth}{!}{%
 \hspace*{-2ex} \includegraphics{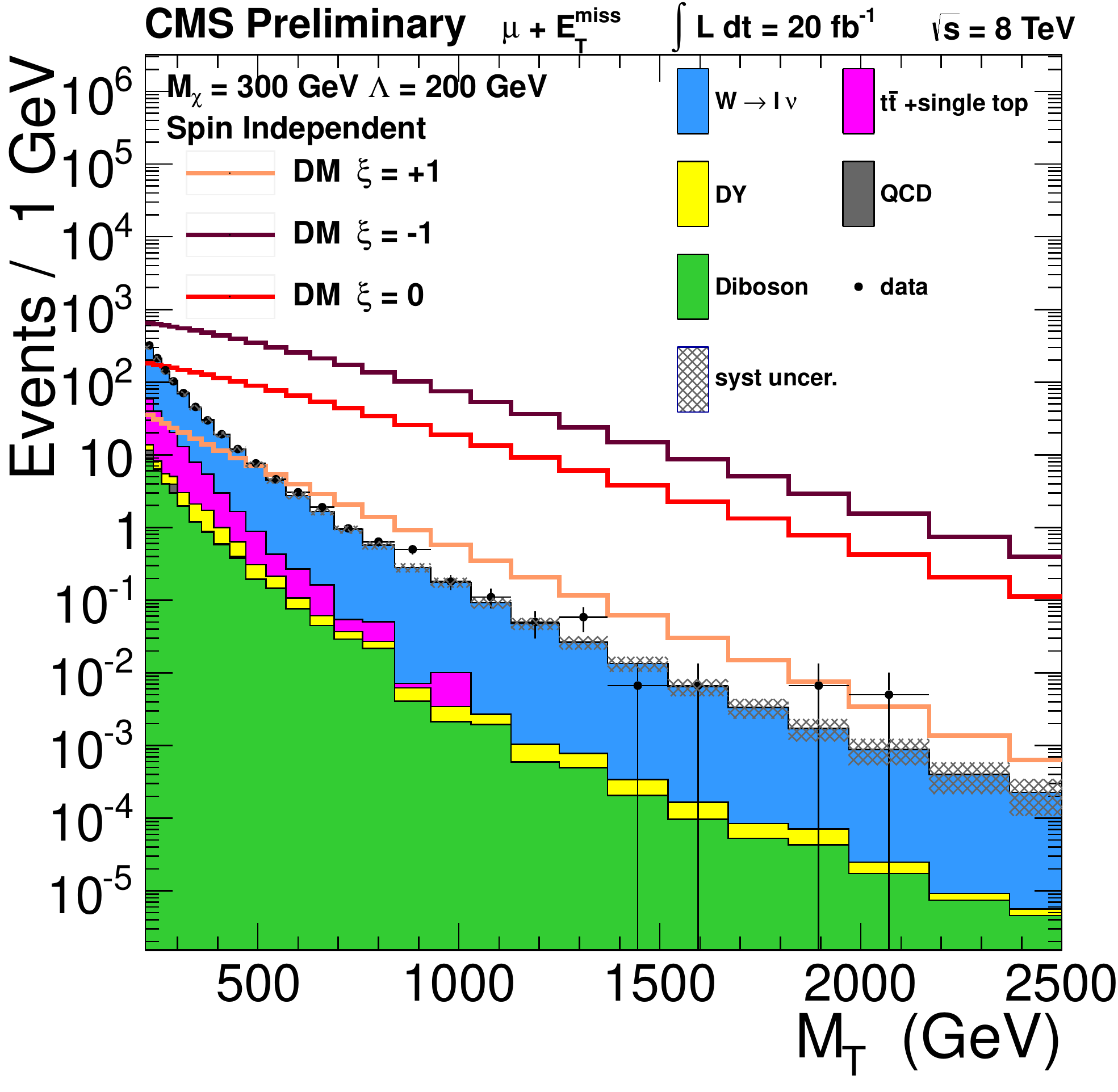} 
 \hspace*{-2ex}
}
\unitlength 1cm
\begin{picture}(10.,6.)
 \put(0.0, 0.0){ \includegraphics[width=7.7cm]{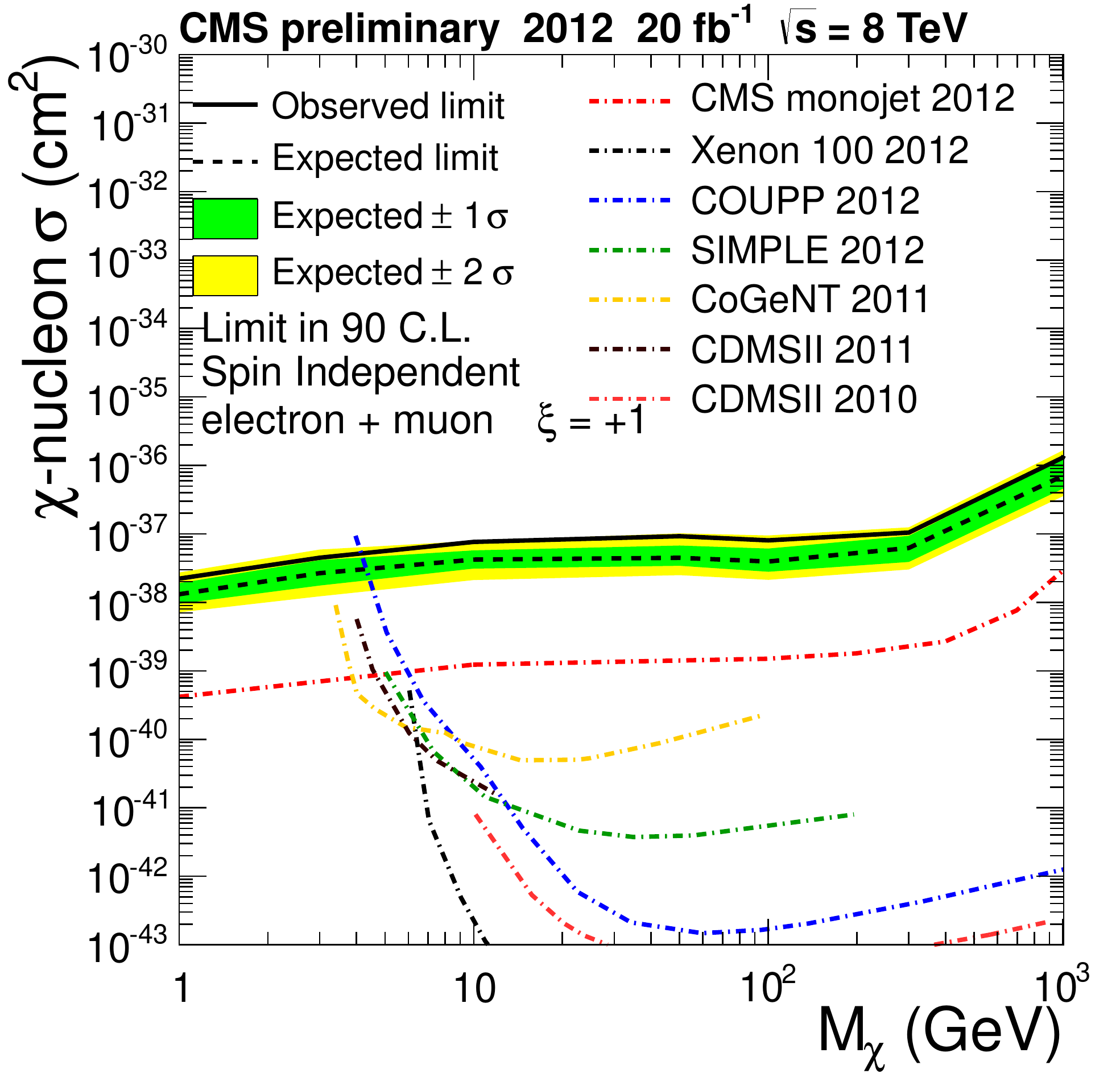} }
\end{picture}

\caption{ \label{DarkMatter}
  Left: Distribution of the transverse mass determined from the lepton and 
$E_T^{\mbox{\scriptsize miss}}$ for the data in good agreement with the SM background predictions.
 Possible Dark Matter (DM) signals with different interference coefficients $\xi$ are superposed.
Right: Limits interpretation of the DM-nucleon cross section as a function of the DM particle 
mass $M_{\chi}$. Limits of dedicated astrophysics experiments and the CMS monojet analysis are
also given. 
}
\end{figure}

\section{Exotic searches for new physics}
Many searches for new physics are pursued by the CMS experiment, ranging from
compositeness, over heavy resonances, long lived particles, lepto-quarks,
fourth generation particles and contact interactions to extra dimensions and black holes.
Limits and their summaries on particle masses and effective energy scales at which new physics 
sets in can be found at \newline
{\tt https://twiki.cern.ch/twiki/bin/view/CMSPublic/PhysicsResultsEXO}. \\
Here a dark matter analysis~\cite{pas-exo-13-004} will be detailed as an example of 
the rich search for new physics programme conducted by the CMS experiment.

A search for new physics in final states with an electron or a muon and $E_T^{\mbox{\scriptsize miss}}$ 
due to particles escaping the detector undetected has been performed. The results can be 
interpreted in terms of the cross section of events with a $W$ boson recoiling against a pair of 
Dark Matter (DM) particles in an effective theory. Such a process would reveal in the 
detector as a lepton plus $E_T^{\mbox{\scriptsize miss}}$ in the final state.
The reconstructed transverse mass distribution of the CMS data in good agreement to SM backgrounds
is shown in Fig.~\ref{DarkMatter} (left).
The DM model considers two possible couplings, vector- and axial-vector
like. They could couple to up- and down-type quarks with different strengths,
parameterised by a coefficient $\xi$. Possible consequences on the signal 
cross section and shape are investigated.
The results are presented in terms of production cross section limits, which are
transformed to limits on the effective parameter $\Lambda$ and $M_{\chi}$-proton cross sections
as measured directly by dedicated astrophysics experiments. 
Figure~\ref{DarkMatter} (right) shows the limits on the $\chi$-nucleon cross section as a 
function of the DM mass $M_{\chi}$. The limits derived on
$\Lambda$ are $\Lambda < 1000/700/300$ GeV for $\xi = -1/ 0/ +1$, for both vector and 
axial-vector couplings.

\begin{figure}[t]
\unitlength 1cm
\begin{picture}(10.,6.)
 \put(-0.2, 0.0){ \includegraphics[width=15.5cm]{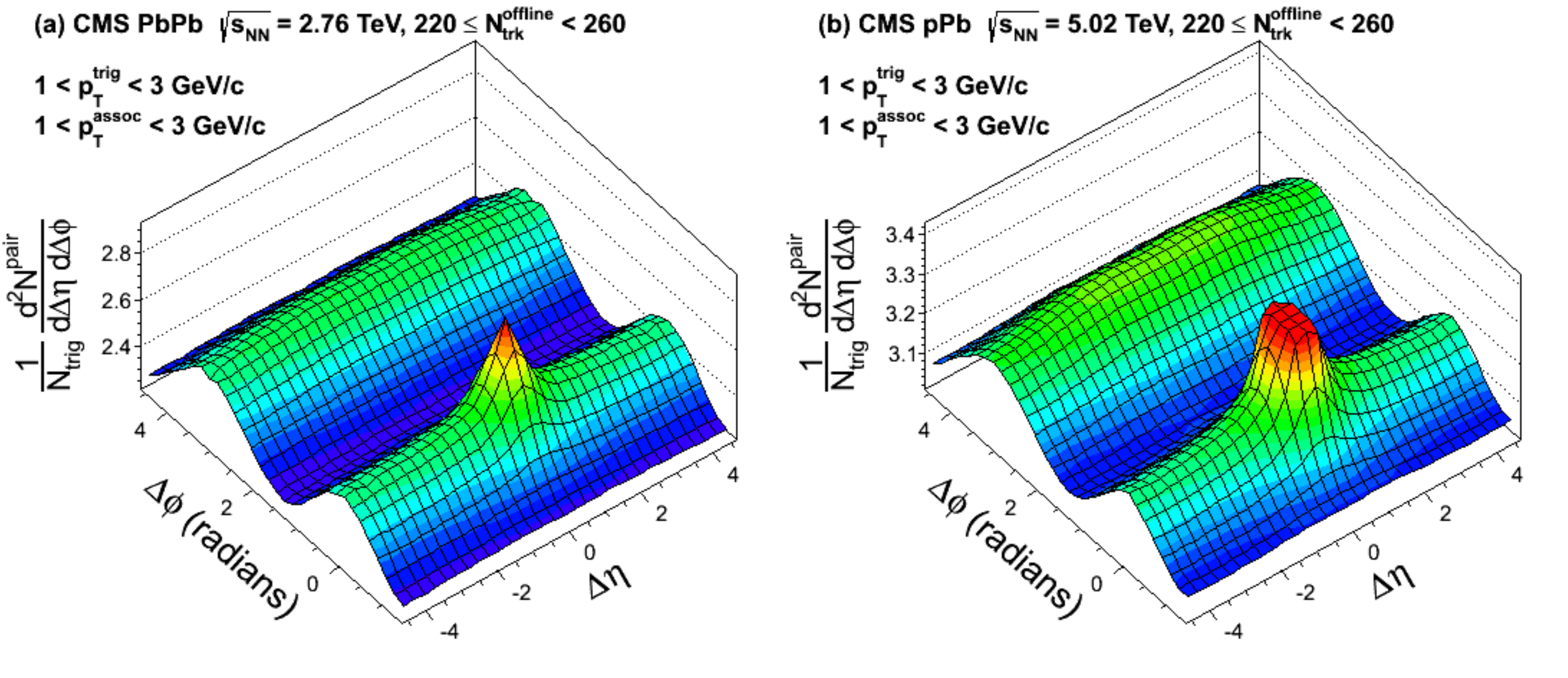} }
\end{picture}

\caption{ \label{HIridge}
2D two-particle correlation functions for 2.76 TeV $Pb-Pb$ (left) and 5.02 TeV $p-Pb$
(right) collisions for pairs of charged particles with $1 < p_T^{\mbox{\scriptsize trig}} < 3$ GeV
and $1 < p_T^{\mbox{\scriptsize assoc.}} < 3$ GeV within the 
$220 \leq N^{\mbox{\scriptsize offline}}_{\mbox{\scriptsize trk}} < 260$ multiplicity bin. 
The sharp near-side peak from jet correlations is truncated to emphasize the structure 
outside that region.
}
\end{figure}

\section{Heavy ion physics}

Studies of multi-particle correlations play a major role in the characterisation of the 
underlying mechanism of particle production in high energy collisions of protons with nuclei. 
Of particular interest in relativistic nucleus-nucleus ($AA$) collisions is the observed 
long-range (large $\Delta |\eta|$) structure in two-dimensional (2D) 
$\Delta\eta$ - $\Delta\phi$ correlation functions. A source of such long-range
correlations is the elliptic flow, induced by the hydrodynamic evolution of the 
collision zone in non-central nucleus-nucleus interactions.
A dedicated high-multiplicity trigger was implemented making use of at least one 
charged particle track with $p_T>0.4$ GeV in the pixel tracker.
The observed long-range correlations can be attributed to the collective expansion of a 
strongly-interacting medium.
In $p-Pb$ collisions, the overall strength of the near-side
correlation is found to be significantly greater than in $pp$ collisions.
The two-particle correlations are shown in Fig.~\ref{HIridge} for $Pb-Pb$ (left)
and $p-Pb$ collisions for pairs of charged particles with $1 < p_T^{\mbox{\scriptsize trig}} < 3$ GeV
and $1 < p_T^{\mbox{\scriptsize assoc.}} < 3$ GeV within the 
$220 \leq N^{\mbox{\scriptsize offline}}_{\mbox{\scriptsize trk}} < 260$ track multiplicity range.
For $Pb-Pb$ collisions, this $N^{\mbox{\scriptsize offline}}_{\mbox{\scriptsize trk}}$ range corresponds to 
an average centrality of approximately 60\%.
In addition to the correlation peak near ($\Delta\eta, \Delta\phi$) = ($0, 0$)
due to jet fragmentation a pronounced long-range structure is seen at
$\Delta\phi \simeq 0$ extending at least 4.8 units in $|\Delta\eta|$.
On the away side ($\Delta\phi \simeq\pi$) of the correlation functions, 
a long-range structure is also seen and found to exhibit a magnitude similar to that 
on the near side for this $p_T$ range.
In non-central $AA$ collisions this $\cos(2\Delta\phi)$-like azimuthal correlation structure 
is assumed to arise primarily from elliptic flow. 
The away-side correlations must also contain contributions from back-to-back jets, which need 
to be accounted for, before extracting any other source of correlations.

\section*{Conclusions}
The LHC accelerator and the CMS experiment provide an excellent performance.
Maintenance, upgrades and optimisations in the present Long Shutdown 1 (LS1) phase are on track.
Forward physics has been accomplished at the LHC at small proton momentum fractions and 
forward rapidities unprecedented at previous colliders.
The Standard Model (SM) has been confirmed, precision measurements have been refined.
A Higgs-like boson has been found and properties are consistent with the assumption of a 
SM Higgs boson. The highest collider energy frontier is being explored.
The decay $B_s\rightarrow \mu\mu$ has been observed (CMS + LHCb combination).
Its Branching ratio is consistent with the SM expectation, putting severe constraints on new 
physics, in particular rendering the Constrained Minimal Supersymmetric SM (CMSSM) unnatural, 
i.e. requiring fine-tuning.
The search for evidence of new physics has been accomplished in many channels.
Long-range correlations in proton-nucleon and nucleon-nucleon collisions provide important
studies on the way for an understanding of the quark-gluon plasma.
More LHC data will be taken at $\sqrt{s} = 13$ TeV after the LS1, ending in spring 2015.

\end{document}